\documentclass[aps,twocolumn,prx,superscriptaddress,amsmath,amssymb]{revtex4-1}
\newcommand{\bras}[1]{\langle#1\rvert}
\newcommand{\kets}[1]{\lvert#1\rangle}

\newcommand{\mean}[1]{\left<#1\right>}
\newcommand{\means}[1]{\langle#1\rangle}

\ifx\pdfoutput\undefined
\usepackage[dvipdfmx]{graphicx}
\usepackage[dvipdfmx]{hyperref}
\else
\usepackage{graphicx}
\usepackage{hyperref}
\fi

\usepackage{txfonts}
\usepackage[T1]{fontenc}
\usepackage{xspace}

\usepackage{dcolumn}
\usepackage{bm}
\usepackage{amsmath}
\usepackage{booktabs}
\usepackage[usenames]{color}
\usepackage{xcolor}

\begin{document}
\let\emph\textit

\title{
Nonequilibrium Majorana Dynamics by Quenching a Magnetic Field in Kitaev Spin Liquids
}

\author{Joji Nasu}
\affiliation{
  Department of Physics, Yokohama National University,
  Hodogaya, Yokohama 240-8501, Japan
}
\author{Yukitoshi Motome}
\affiliation{
  Department of Applied Physics, University of Tokyo,
  Bunkyo, Tokyo 113-8656, Japan
}

\date{\today}
\begin{abstract}
The honeycomb Kitaev spin model provides a quantum spin liquid in the ground state, where the spin excitations are fractionalized into itinerant and localized Majorana fermions;
the former spectrum has a broad continuum ranging up to a high energy, while the latter has a sharp  peak at a low energy.
Despite tremendous efforts, it remains elusive to clearly identify these distinct Majorana excitations in experiments.
Here we show their manifestation in the time evolution after quenching the magnetic field, by using the time-dependent Majorana mean-field theory for both the ferromagnetic and antiferromagnetic Kitaev models.
We find that the transient spin dynamics from the quantum spin liquid states is qualitatively different from the conventional spin precessions by the quench from the high-field forced-ferromagnetic state.
We obtain peculiar time evolutions with distinct time scales, i.e., short-time decay of high-energy components associated with the itinerant Majorana excitations, and long-lived excitations at a low energy by the localized ones.
These peculiar behaviors are caused by the energy transfer between the two Majorana quasiparticles after the field quench.
Moreover, we find that the Majorana semimetal with the point nodes in equilibrium turns into a Majorana metal with the transient ``Fermi surfaces'' by the energy transfer.
In particular, for the quench from the intermediate-field quantum spin liquid in the antiferromagnetic Kitaev model, the Fermi surfaces change their topology in the time evolution, which is regarded as a dynamical version of the Majorana ``Lifshitz transition''.
Our results unveil that the real-time dynamics provides another route to not only the identification of the fractional Majorana excitations in candidate materials of Kitaev magnets but also unprecedented quantum phases that cannot be stabilized as the equilibrium states.
\end{abstract}

\maketitle

\section{Introduction}

Fractionalization of fundamental degrees of freedom in quantum many-body states is one of the most fascinating subjects in condensed matter physics.
Amongst others, quantum spin liquids (QSLs), where any long-range magnetic order is absent down to the lowest temperature, have been intensively studied as a playground for the fractionalization of the spin degree of freedom~\cite{ISI:000275366100033,Savary2017}.
Antiferromagnets on geometrically frustrated lattices are considered as typical candidates exhibiting the QSLs~\cite{Anderson1973153}.
In the QSLs, elementary excitations from the ground state are predicted to be described as fractional quasiparticles, such as spinons and visons~\cite{Read1989,Read1991,Wen1991,Baskaran1988,Senthil2000,PhysRevLett.96.060601}.
However, the clear identification of such fractional excitations, especially in more than one dimension, is a long-standing challenge in both theories and experiments.

Recently, the Kitaev model, which is a quantum spin model with bond-dependent interactions between localized $S=1/2$ magnetic moments, has attracted considerable attention as its ground state is exactly shown to be a QSL~\cite{Kitaev2006,PhysRevLett.98.247201,Trebst2017pre,Hermanns2018rev,Knolle2019rev,takagi2019rev}.
The excitations from the QSL ground state are described by two kinds of Majorana quasiparticles emergent from the fractionalization of spins:
One is itinerant Majorana fermions, whose energy band ranges continuously from zero energy to a high energy of $O(J)$ ($J$ is the bare exchange coupling), and the other is localized Majorana fermions, whose excitations are gapped and form a flat band at a low energy of $O(0.1J)$.
To identify these fractional quasiparticles with distinct energy scales, thermodynamic quantities, spin dynamics, and transport properties have been  investigated theoretically~\cite{PhysRevLett.102.017205,PhysRevLett.105.027204,PhysRevLett.112.207203,Jiang2011,PhysRevLett.113.107201,PhysRevB.92.115127,PhysRevB.92.115122,PhysRevLett.113.187201,yoshitake2016,Song2016,Nasu2016nphys,Winter2016,yadav2016kitaev,Yoshitake2017PRBa,Yoshitake2017PRBb,winter2017breakdown,Gohlke2017,Nasu2017,PhysRevB.93.174425,catuneanu2018path,Udagawa2018,Ronquillo2019} and measured in candidate materials such as iridium oxides and ruthenium compounds~\cite{PhysRevB.82.064412,PhysRevLett.108.127203,PhysRevB.90.041112,PhysRevB.91.094422,PhysRevB.91.144420,Majumder2015,Kitagawa2018nature,PhysRevLett.114.147201,banerjee2016proximate,Banerjee2017,hirobe2017,Kasahara2018,kasahara2018majorana}.
Despite tremendous attempts, the experimental observation of these quasiparticles remains elusive.
This is partly because the spin excitations measured in experiments are usually given by a composite of the two types of Majorana fermions, and it is hard to observe them separately.

An effective technique to observe elementary excitations with distinct energy scales is nonequilibrium measurements for transient dynamics.
For instance, in Mott insulators in low dimensions, the transient carrier injection via photoirradiation was theoretically examined for the observation of spin and change excitations separately~\cite{Takahashi2002,yonemitsu2008theory,Oka2008}.
Indeed, the spin-charge separation was clarified experimentally in quasi-one-dimensional organic compounds by using the femtosecond pump-probe spectroscopy~\cite{iwai2003,okamoto2006,okamoto2007}.
In addition, for quasi-two-dimensional organic molecular salts exhibiting a charge disproportionation, pump-probe experiments clarified that two excitations, high-energy charge dynamics and low-energy molecular vibrations, appear to be decoupled in the time domain; the charge dynamics is observed in the early-time stage and quickly decays, and after that, the molecular vibration dynamics is observed for a longer time~\cite{Iwai2007,Nakaya2010,Kawakami2010}.
Thus, the transient dynamics is useful to identify distinct excitations in strongly correlated electron systems.

A similar technique for nonequilibrium dynamics might be a promising tool to observe the fractional quasiparticles with distinct energy scales in the Kitaev QSLs.
Thus far, several theoretical works have been done for the nonequilibrium spin dynamics of the Kitaev model~\cite{Sengupta2008,Mondal2008,Hikichi2010,Patel2012,Sato2014pre,bhattacharya2016,Rademaker2017pre,Sameti2019}.
For instance, time evolutions by quenching the exchange interactions were discussed, and unconventional scaling laws were found for correlation functions~\cite{Mondal2008,Hikichi2010,Patel2012}.
Moreover, the Floquet states under periodically-altered exchange interactions were studied to realize the unusual change of the Majorana fermion bands, such as nonequilibrium topological transitions~\cite{Sato2014pre,bhattacharya2016}.
Although these theoretical proposals are intriguing, their experimental confirmation has not yet been done, mainly because it is difficult to realize such a real-time modulation of the exchange interactions experimentally.
Furthermore, it is not obvious how the fractional excitations are identified in the dynamical behaviors.
It is therefore desired to propose another feasible way of driving interesting nonequilibrium dynamics related with the fractional excitations in the Kitaev QSLs.

In this paper, we investigate the transient spin dynamics in the Kitaev model yielded by quenching an external magnetic field.
By exploiting a time-dependent version of the mean-field (MF) approximation in the Majorana fermion representation (Majorana MF theory)~\cite{Nasu2018mag}, we calculate the time evolutions of the magnetization and spin correlations after the field quench for both ferromagnetic (FM) and antiferromagnetic (AFM) Kitaev models.
We find that the transient dynamics is qualitatively different depending on the quantum phase before quenching.
In the case of the quench from the high-field forced-FM phase, where the fractionalization is absent, the transient spin dynamics is a conventional one originating from the precessional motion of spins; two Majorana fermions are strongly coupled and indistinguishable in the time evolution.
In stark contrast, in the quench from the low-field QSL phase connected to the zero-field Kitaev QSL, two quasiparticles are separately observed in the time evolutions of spin correlations with distinct time scales.
We show that such fractional dynamics is observed in more peculiar manner for the quench from the intermediate-field QSL phase that appears only in the AFM Kitaev model.
The peculiar behaviors are discussed by the transfer of the exchange energy from the itinerant to localized Majorana quasiparticles by the field quench.
We also find that the energy transfer turns the Majorana semimetal with the point nodes into a Majorana metal with the transient ``Fermi surfaces'' in the Majorana fermion bands.
In particular, in the case of the intermediate-field QSL in the AFM Kitaev model, one of the Majorana Fermi surfaces evolves from open to closed one, which is regarded as a dynamical version of the ``Lifshitz transition''.
We discuss the possibility of experimental observations of our results, e.g., by magneto-optical effects and transient hidden phases via the Peierls instability.

This paper is organized as follows.
In Sec.~\ref{sec:model}, we introduce the Kitaev model with a time-dependent magnetic field and its Majorana fermion representation.
We also discuss the fundamental aspect of the spin fractionalization and its implication in the dynamical spin structure factor, which is relevant in the following sections.
In Sec.~\ref{sec:method}, we present the framework of the time-dependent Majorana MF theory.
In Sec.~\ref{sec:result}, we show the results for the transient dynamics of the magnetization and spin correlations (Sec.~\ref{sec:time-evol-magn}), and the Majorana fermion states (Sec.~\ref{sec:time-evol-major} and \ref{sec:time-evol-majorDOS}).
We discuss the relevance of our results in Sec.~\ref{sec:discussion}, focusing on the experimental observations.
Finally, Sec.~\ref{sec:summary} is devoted to the summary.
The validity of the present method is discussed in Appendix~\ref{sec:valid-time-depend}.

\section{Model}\label{sec:model}

\begin{figure}[t]
 \begin{center}
  \includegraphics[width=0.8\columnwidth,clip]{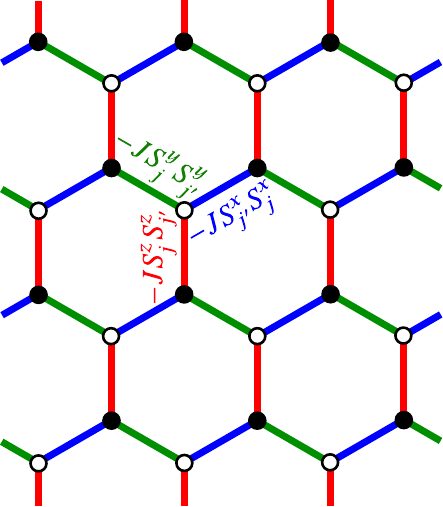}
  \caption{
Schematic picture of the Kitaev model in Eq.~\eqref{eq:1} on the honeycomb structure.
The bond-dependent Ising-type interaction $-J S_j^\gamma S_{j'}^\gamma$ ($\gamma=x,y,z$) is defined on the $\gamma$ bonds represented by blue, green, and red for $\gamma=x$, $y$, and $z$, respectively.
}
  \label{fig_lattice}
 \end{center}
\end{figure}

We consider the Kitaev model under the time-dependent magnetic field, whose Hamiltonian is given by
\begin{align}
 {\cal H}(t)=-J\sum_{\means{jj'}_\gamma}S_j^\gamma S_{j'}^\gamma -h(t) \sum_j S_j^z,\label{eq:1}
\end{align}
where $\means{jj'}_\gamma$ stands for a nearest-neighbor (NN) bond of the honeycomb lattice connecting sites with $S=1/2$ localized spins and the superscript $\gamma(=x,y,z)$ distinguishes three kinds of inequivalent bonds shown in Fig.~\ref{fig_lattice}.
We assume that all exchange constants are the same as $J$, and $h(t)$ is the magnetic field along the $S^z$ direction as a function of time $t$.
While the following calculations can be applied to any $h(t)$, we here limit ourselves to the time dependence given by
\begin{align}
 h(t)=
\begin{cases}
 h & {\rm for} \ \ t\leq0\\
 0 & {\rm for} \ \ t> 0
\end{cases},\label{eq:4}
\end{align}
which mimics a sudden quench of the magnetic field.

By applying the Jordan-Wigner transformation and introducing a Majorana fermion representation~\cite{PhysRevB.76.193101,PhysRevLett.98.087204,1751-8121-41-7-075001,PhysRevLett.113.197205,PhysRevB.92.115122} to Eq.~(\ref{eq:1}), we can rewrite the spin model in Eq.~\eqref{eq:1} into the interacting Majorana fermion model as
\begin{align}
 {\cal H}(t)=&-\frac{iJ}{4}\sum_{\gamma=x,y}\sum_{\mean{jj'}_\gamma}a_j b_{j'}-\frac{J}{4}\sum_{\mean{jj'}_z}i a_j b_{j'}i \bar{a}_j \bar{b}_{j'}\nonumber\\
&-\frac{ih(t)}{2}\left(\sum_{j\in A}a_j\bar{a}_j-\sum_{j'\in B}b_{j'}\bar{b}_{j'}\right),
 \label{eq:H_Maj_S}
\end{align}
where $a_j$ and $\bar{a}_j$ ($b_{j'}$ and $\bar{b}_{j'}$) are the Majorana fermions defined on site $j$ ($j'$) of the $A$ ($B$) sublattice, which is shown by the black (white) circles in Fig.~\ref{fig_lattice}.
The spin operators on the $A$ and $B$ sublattices are explicitly given by
\begin{align}
 S_j^x=\frac{a_j}{2}\prod_{j''<j}\left(-2S_{j''}^z\right), \
 S_j^y=-\frac{\bar{a}_j}{2}\prod_{j''<j}\left(-2S_{j''}^z\right), \
 S_j^z=\frac{i}{2}a_j\bar{a}_j,
\end{align}
and
\begin{align}
S_{j'}^x=\frac{\bar{b}_{j'}}{2}\prod_{j''<j'}\left(-2S_{j''}^z\right), \
 S_{j'}^y=-\frac{b_{j'}}{2}\prod_{j''<j'}\left(-2S_{j''}^z\right), \
 S_{j'}^z=\frac{i}{2}\bar{b}_{j'} b_{j'},
\end{align}
respectively, where the sites are numbered along the chain consisting of the $x$ and $y$ bonds.

The Kitaev model is exactly solvable in the absence of the magnetic field ($h=0$).
The ground state is obtained as an eigenstate in the Majorana representation in Eq.~\eqref{eq:H_Maj_S} without the last term;
the Majorana fermions $\{\bar{a}, \bar{b}\}$ are localized on the $z$ bonds, while $\{a, b\}$ are itinerant.
The excitations from the ground state are described by the itinerant and localized Majorana fermions.
These two types of Majorana fermion excitations have distinct energy scales, and affect thermodynamics and spin dynamics in a particular manner.
Among them, the dynamical spin structure factor ${\cal S}(\bm{q},\omega)$, which is measured in neutron scattering experiments, is important for the following discussions in the time evolutions of the magnetization and spin correlations.
Let us focus on the uniform component ${\cal S}(\bm{q}=0,\omega)$, which is relevant in the quench of the spatially uniform magnetic field considered here.
In the FM case ($J>0$), ${\cal S}(\bm{q}=0,\omega)$ exhibits two distinct structures: a low-energy coherent peak at $\sim0.1J$ and high-energy incoherent feature up to $\sim2J$~\cite{PhysRevLett.112.207203,PhysRevB.92.115127}.
The former mainly originates from the localized Majorana fermions $\{\bar{a}, \bar{b}\}$, while the latter from the itinerant ones $\{a, b\}$.
On the other hand, such a coherent peak appears around the Brillouin zone boundary and is absent at $\bm{q}=0$ in the AFM case ($J<0$); ${\cal S}(\bm{q}=0,\omega)$ has only a broad incoherent spectrum~\cite{PhysRevLett.112.207203,PhysRevB.92.115127} (see also Fig.~\ref{fig_sw} in Appendix~\ref{sec:valid-time-depend}).

A nonzero $h$ hybridizes the two types of Majorana fermions through the last term in Eq.~\eqref{eq:H_Maj_S}, and the exact solvability is lost (see Sec.~\ref{sec:band_t=0}).
In the following, we describe the ground state before the field quench ($t\leq 0$) by using the Majorana MF theory, and track the time evolution of the wave function by using the time-dependent version of the Majorana MF theory introduced in the next section.

\section{Method}\label{sec:method}

\subsection{Time-dependent Majorana MF theory}\label{sec:time-dependent-hf}

In this section, we introduce the Majorana MF theory for the time-dependent Hamiltonian.
Before introducing the framework, let us briefly review the Majorana MF theory for the Kitaev model in the equilibrium state under a static magnetic field~\cite{Nasu2018mag}.
In this theory, the Majorana interactions in the second term of Eq.~(\ref{eq:H_Maj_S}) are decoupled by introducing the Hartree-Fock type MFs as
$i a_j b_{j'}i \bar{a}_j \bar{b}_{j'}\simeq -i X b_{j'} \bar{b}_{j'} -iY a_j \bar{a}_j +XY
+i\bar{\Phi} a_j b_{j'} +i\Phi \bar{a}_j \bar{b}_{j'} -\Phi\bar{\Phi}
-i\Theta \bar{a}_j b_{j'} -i\bar{\Theta} a_j \bar{b}_{j'} +\Theta\bar{\Theta}$,
where  $j$ ($j'$) are the $A$($B$)-sublattice site on the corresponding $z$ bond and the MFs are defined by
 $X=i\means{a_j \bar{a}_j}$, $Y=i\means{b_{j'} \bar{b}_{j'}}$, $\Phi=i\means{a_j b_{j'}}$, $\bar{\Phi}=i\means{\bar{a}_j \bar{b}_{j'}}$, $\Theta=i\means{a_j \bar{b}_{j'}}$, and $\bar{\Theta}=i\means{\bar{a}_j b_{j'}}$.
Note that the MFs are assumed to be spatially uniform for simplicity.
Then, the MF Hamiltonian is obtained in the bilinear form in terms of the Majorana fermion operators, which can be easily diagonalized in the reciprocal space.
The details are given in Supplemental Material in Ref.~\cite{Nasu2018mag}.

We extend the Majorana MF theory for the time-dependent Hamiltonian in Eq.~\eqref{eq:H_Maj_S}, following the conventional time-dependent MF theory applied to correlated electron systems~\cite{terai1993solitons,Hirano2000,Tanaka2010,Ohara_2017,Tanaka2018,Seo2018}.
In this framework, each one-particle state $\kets{\phi_{\bm{k}\nu}(t)}$ evolves with time $t$ obeying the Shr\"odinger equation:
\begin{align}
 i\frac{\partial \kets{\phi_{\bm{k}\nu}(t)}}{\partial t}={\cal H}_{\bm{k}}^{\rm MF}(t)\kets{\phi_{\bm{k}\nu}(t)},
\label{eq:Schrodinger_eq}
\end{align}
where ${\cal H}_{\bm{k}}^{\rm MF}(t)$ is the MF Hamiltonian in the reciprocal space with wavenumber $\bm{k}$.
This equation can be formally solved as
\begin{align}
 \kets{\phi_{\bm{k}\nu}(t)}={\cal T}\exp\left[-i\int_0^t {\cal H}_{\bm{k}}^{\rm MF}(t')dt'\right]\kets{\phi_{\bm{k}\nu}(0)},\label{eq:2}
\end{align}
where ${\cal T}$ is the time-ordering operator.
Thus, when considering an infinitesimal time evolution $\delta t$, the wave function is obtained as
\begin{align}
 \kets{\phi_{\bm{k}\nu}(t+\delta t)}=\sum_\mu \means{\varphi_{\bm{k}\mu}(t) | \phi_{\bm{k}\nu}(t)} e^{-i\varepsilon_{\bm{k}\mu}(t) \delta t}\kets{\varphi_{\bm{k}\mu}(t)},\label{eq:3}
\end{align}
where $\varphi_{\bm{k}\mu}(t)$ is a one-particle eigenstate of ${\cal H}_{\bm{k}}^{\rm MF}(t)$ with the eigenenergy $\varepsilon_{\bm{k}\mu}(t)$.
Note that the norm of $\kets{\varphi_{\bm{k}\mu}(t)}$ is conserved and taken to be unity.

In the present case, we start from the system with a nonzero magnetic field $h>0$ at $t=0$.
We describe the initial state by using the Majorana MF theory for the equilibrium state; the MFs are obtained by performing iterative calculations until the convergence, and the one-particle occupied eigenstates $\kets{\phi_{\bm{k}\nu}(0)}$ are calculated for the converged MFs.
At $t=0^+$, the magnetic field is quenched to zero, and $\kets{\phi_{\bm{k}\nu}(0)}$ is no longer the eigenstate of the MF Hamiltonian ${\cal H}_{\bm{k}}^{\rm MF}(t>0)$.
We diagonalize ${\cal H}_{\bm{k}}^{\rm MF}(0^+)$ by using the MFs at $t=0$, and obtain $\varphi_{\bm{k}\mu}(0^+)$ and $\varepsilon_{\bm{k}\mu}(0^+)$.
By applying Eq.~(\ref{eq:3}) for a small time evolution $\Delta t$, we can evaluate $\kets{\phi_{\bm{k}\nu}(\Delta t)}$.
Using the many-body state composed of the occupied states $\kets{\phi_{\bm{k}\nu}(\Delta t)}$, we calculate the MFs at $t=\Delta t$ and construct the MF Hamiltonian ${\cal H}_{\bm{k}}^{\rm MF}(\Delta t)$ with these MFs.
Then, we can evaluate $\kets{\phi_{\bm{k}\nu}(2\Delta t)}$ by using Eq.~(\ref{eq:3}).
By repeating the above procedures, we can compute the time evolutions of the wave function, the eigenenergy, and the MFs.
In the following calculations, the time step is taken to be $\Delta t /|J|^{-1}=0.00133$.
The validity of the present method and the numerical precision are discussed in Appendix~\ref{sec:valid-time-depend}.

\subsection{Wavelet analysis}\label{sec:wavelet-analysis}

To analyze the time dependence of physical quantities, we adopt the continuous wavelet transformation, which is widely used for time-dependent spectra.
The wavelet transformation of the time-dependent function $f(t)$ is generally given by
\begin{align}
 w(a,b)=\frac{1}{\sqrt{a}}\int  \left\{ f(t')-f_0 \right\}  \left\{\psi\left(\frac{t'-b}{a}\right)\right\}^* dt',
\label{eq:wavelet}
\end{align}
where $a$ and $b$ are the scaling and time parameters, respectively, and $f_0$ is the long-time average of $f(t)$.
In the present study, we use the Morlet wavelet for $\psi(t)$ in Eq.~\eqref{eq:wavelet}, which is given by
\begin{align}
 \psi(t)=\pi^{-1/4 }\left(e^{i\omega_0 t}-e^{-\omega_0^2/2}\right)e^{-t^2/2},
\end{align}
where $\omega_0$ is the dimensionless center frequency.
The wavelet scalogram $W(t,\omega)$ is obtained by using Eq.~\eqref{eq:wavelet} as
\begin{align}
W(t,\omega)=\left|w\left(\frac{\omega_0}{\omega},t\right)\right|.
\end{align}
In the following calculations, we take the long-time average $f_0$ for $60\leq t/|J|^{-1}\leq266.7$ and $\omega_0=6$.

\section{Result}\label{sec:result}

\subsection{Time evolution of magnetization and spin correlations}\label{sec:time-evol-magn}

\subsubsection{Initial state at $t=0$}\label{sec:result_h=0}

\begin{figure*}[t]
 \begin{center}
  \includegraphics[width=2\columnwidth,clip]{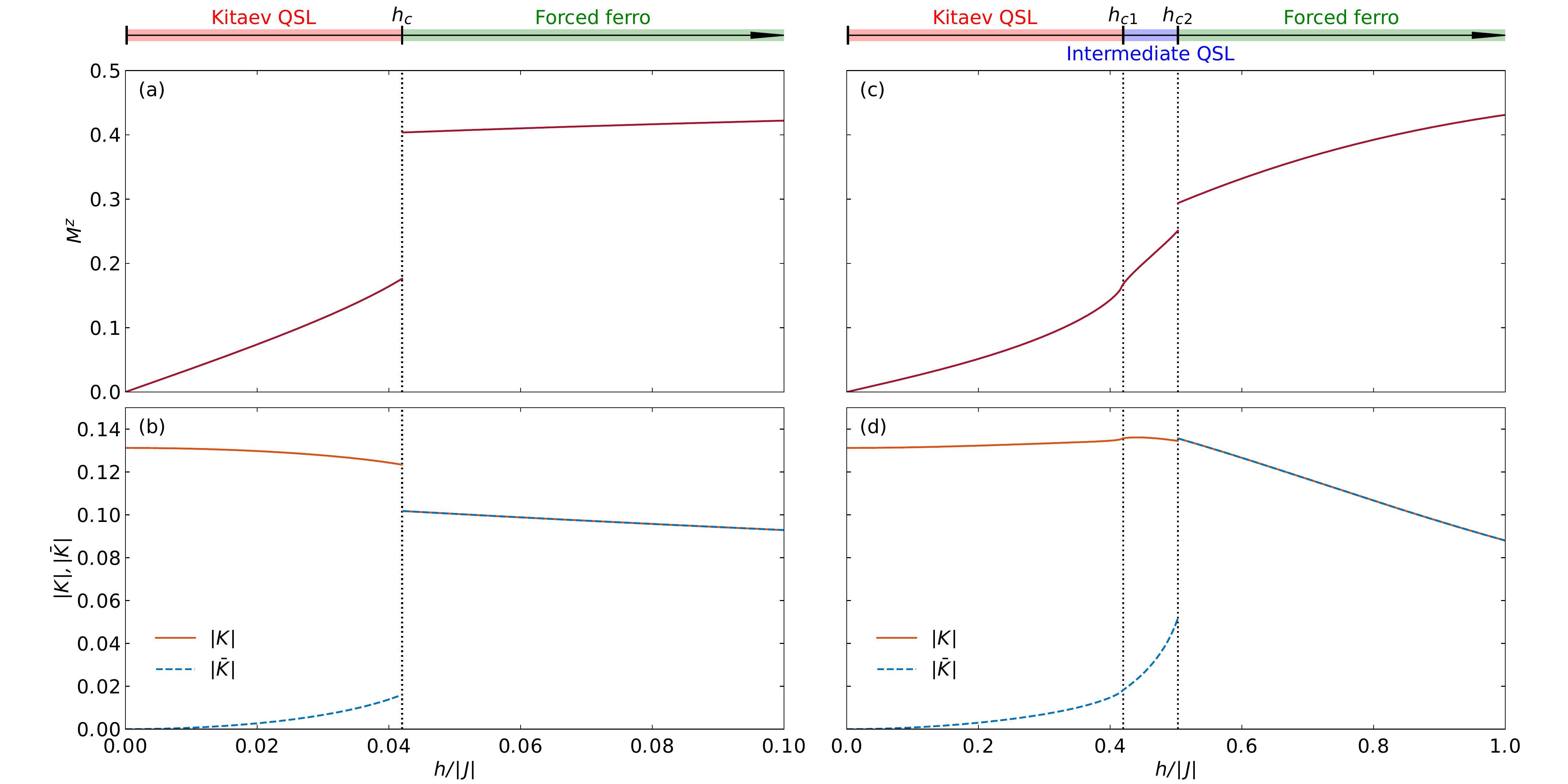}
  \caption{
Magnetic-field dependence of (a),(c) the magnetization and (b),(d) the kinetic energy of the Majorana fermions for (a),(b) the FM Kitaev model and (c),(d) the AFM one under the static magnetic field.
}
  \label{fig_hdep_full}
 \end{center}
\end{figure*}

Before showing the time evolution after the field quench, let us discuss the initial state at $t=0$ under the magnetic field $h$.
As described in Sec.~\ref{sec:time-dependent-hf}, the initial state is obtained by the static Majorana MF theory; the results were obtained in the previous study for both FM and AFM cases~\cite{Nasu2018mag}.
Figure~\ref{fig_hdep_full}(a) and~\ref{fig_hdep_full}(c) show the magnetization in the $z$ direction, $M^z=\frac{1}{N}\sum_j\means{S_j^z}$, as a function of the static field $h$ for the FM and AFM Kitaev models, respectively ($N$ is the number of spins in the system).
In the FM case, there is a discontinuous phase transition at $h_c/|J|\simeq 0.0421$ associated with a jump of the magnetization as shown in Fig.~\ref{fig_hdep_full}(a).
This is a phase transition from the low-field Kitaev QSL phase to the high-field forced-FM phase [see the phase diagram above Fig.~\ref{fig_hdep_full}(a)].
On the other hand, in the AFM case, two successive phase transitions take place: continuous one at $h_{c1}/|J|\simeq 0.417$ and discontinuous one at $h_{c2}/|J|\simeq 0.503$, as shown in Fig.~\ref{fig_hdep_full}(c).
Thus, the AFM Kitaev model shows an intermediate phase between the low-field Kitaev QSL and the high-field forced-FM phases [see the phase diagram above Fig.~\ref{fig_hdep_full}(c)].
This intermediate phase is identified as another QSL, and the continuous phase transition at $h_{c1}$ is accompanied by a topological change in the Majorana fermion band~\cite{Nasu2018mag,Liang2018}.
A similar intermediate QSL phase in the AFM case was pointed out also by other theoretical calculations~\cite{Zhu2018,Gohlke2018,hickey2019emergence,Patel2018pre}.

As mentioned in Sec.~\ref{sec:model}, at $h=0$, the spin excitations are fractionalized into itinerant and localized Majorana fermions, $\{a, b\}$ and $\{\bar{a}, \bar{b}\}$, respectively.
The introduction of $h$ hybridizes these two quasiparticles through the last term in Eq.~\eqref{eq:H_Maj_S}, which makes $\{\bar{a}, \bar{b}\}$ also itinerant in the presence of the magnetic field (see the discussion of the Majorana band structure in Sec.~{\ref{sec:band_t=0}).
To see these behaviors, we introduce the kinetic energies of the two kinds of Majorana fermions as
\begin{align}
 K=-\frac{1}{2N}\sum_{\means{jj'}_x}\means{i a_j b_{j'}}, \
\bar{K}=-\frac{1}{2N}\sum_{\means{jj'}_y}\means{i \bar{a}_j \bar{b}_{j'}}.
\end{align}
In the original spin representation, these are equivalent to the spin correlations:
\begin{align}
 K=\frac{2}{N}\sum_{\means{jj'}_x}\means{S_j^x S_{j'}^x}, \
\bar{K}=\frac{2}{N}\sum_{\means{jj'}_y}\means{S_j^x S_{j'}^x}.
\label{eq:7}
\end{align}
Note that $K$ describes the spin correlation of the $x$ component on the $x$ bonds, namely, for the spin component connected by the exchange coupling $J$, whereas $\bar{K}$ is for the $x$ component on the $y$ bonds, namely, the noninteracting spin component [see Eq.~\eqref{eq:1}].
Figures~\ref{fig_hdep_full}(b) and~\ref{fig_hdep_full}(d) show $|K|$ and $|\bar{K}|$, respectively, as functions of $h$.
In the absence of the magnetic field, $|\bar{K}|$ is zero while $|K|$ is nonzero in both FM and AFM cases, indicating that the Majorana fermions $\{\bar{a},\bar{b}\}$ are localized while $\{a,b\}$ are itinerant.
By introducing $h$, $|\bar{K}|$ becomes nonzero because of the hybridization as expected above.
In the forced-FM phase, the two kinetic energies take the same value in both FM and AFM cases, which reflects the disappearance of the spin fractionalization.

\begin{figure*}[t]
 \begin{center}
  \includegraphics[width=2\columnwidth,clip]{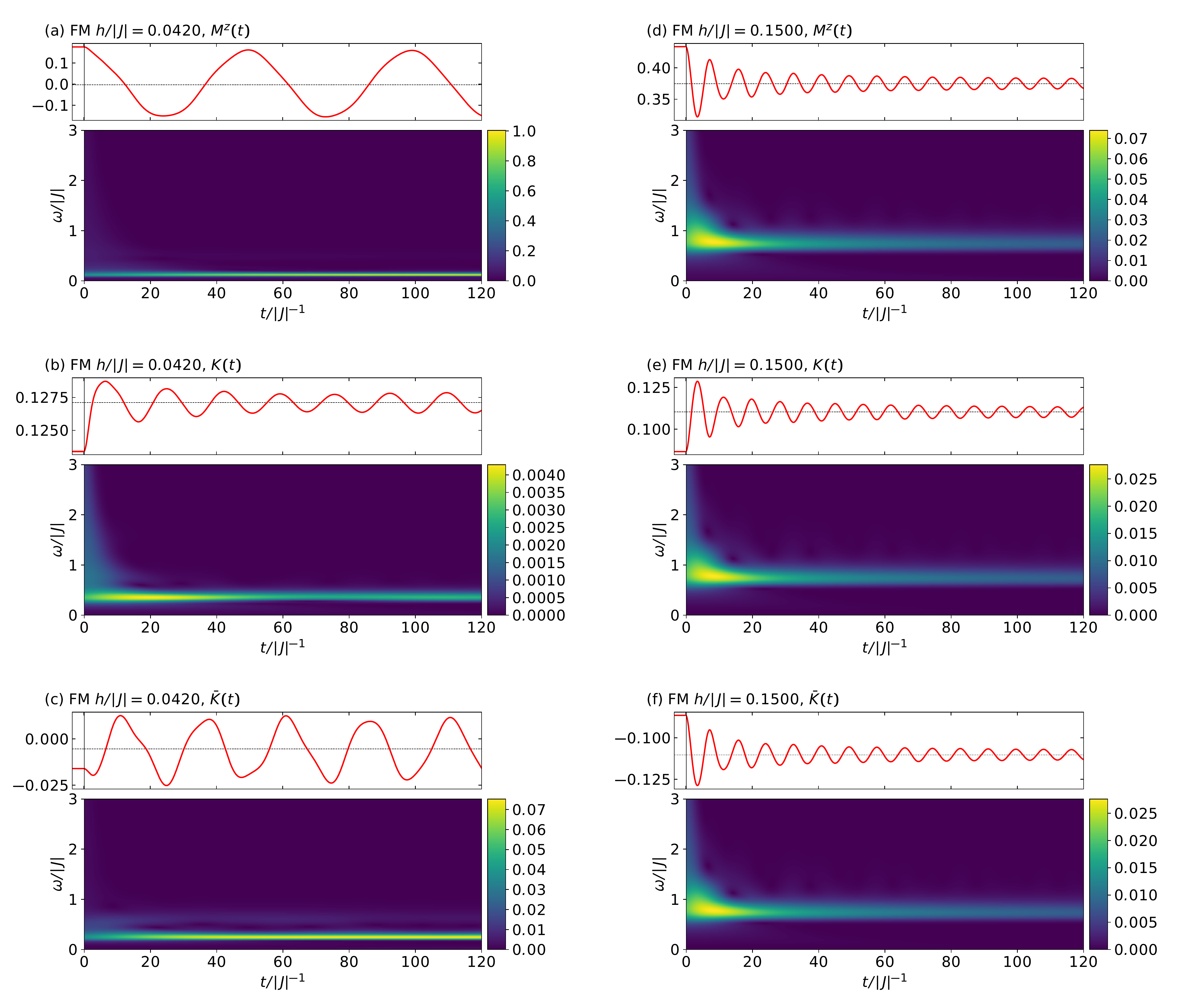}
  \caption{
Time evolutions of (a) $M^z$, (b) $K$, and (c) $\bar{K}$, and their wavelet scalograms in the FM Kitaev model for the field quench from $h/J=0.0420$ (Kitaev QSL state).
(d)--(f) Corresponding plots from $h/J=0.15$ (forced-FM state).
The dotted lines represent the long-time averages of each quantity.
}
  \label{fig_waveletF}
 \end{center}
\end{figure*}

\subsubsection{Ferromagnetic case}
\label{sec:time-evol-magn_ferro}

First, we show the results of the time evolution in the FM Kitaev model.
The upper panel of Fig.~\ref{fig_waveletF}(a) shows the time dependence of the magnetization $M^z(t)$ after the field quench from the QSL state at $h/|J|=0.0420$ just below $h_c/|J|\simeq0.0421$.
The magnetization oscillates around zero after the magnetic field vanishes.
The time dependence is close to a simple cosine curve, implying the weak time dependence of the frequency.
This is confirmed by the wavelet transformation introduced in Sec.~\ref{sec:wavelet-analysis}, as shown in the wavelet scalogram in the lower panel of Fig.~\ref{fig_waveletF}(a); the frequency is almost constant $\sim 0.1|J|$ as a function of time.
This energy scale is close to that of the sharp peak in the dynamical spin structure factor ${\cal S}(\bm{q}=0,\omega)$ in the equilibrium state at $h=0$, which predominantly originates from the Majorana fermions $\{\bar{a},\bar{b}\}$, as discussed in Sec.~\ref{sec:model}~\cite{PhysRevLett.112.207203,PhysRevB.92.115127} [see also Fig.~\ref{fig_sw}(a) and Eq.~(\ref{eq:5}) in Appendix~\ref{sec:valid-time-depend}].
Thus, the result suggests that the long-lived slow oscillation of the magnetization predominantly governed by the Majorana excitations described by $\{\bar{a},\bar{b}\}$.

We also calculate the time developments of $K$ and $\bar{K}$.
The results are shown in Figs.~\ref{fig_waveletF}(b) and~\ref{fig_waveletF}(c) with their wavelet scalograms.
In the early-time period up to $t/|J|^{-1}\sim 5$, there is a high-energy broad structure with a relatively weak spectral weight in $K$, as shown in the scalogram of Fig.~\ref{fig_waveletF}(b).
This appears to correspond to the high-energy incoherent spectrum in ${\cal S}(\bm{q}=0,\omega)$~\cite{PhysRevLett.112.207203,PhysRevB.92.115127} [see Fig.~\ref{fig_sw}(a) in Appendix~\ref{sec:valid-time-depend}].
While increasing $t$, this broad structure disappears and there remains a long-lived oscillation with the frequency of $\sim 0.4|J|$.
On the other hand, $\bar{K}$ does not show such a high-energy broad structure, while the oscillation is somewhat deformed in the early stage, as shown in Fig.~\ref{fig_waveletF}(c).
The frequency of the long-lived component appears at $\sim 0.25|J|$ and shows different time evolution from that of $K$; while the intensity of $K$ is gradually suppressed by the elapse of time, that of $\bar{K}$ is enhanced.
The distinct time dependence between the Majorana fermions $\{a,b\}$ and $\{\bar{a},\bar{b}\}$ is interpreted as a consequence of the spin fractionalization observed in the time domain.
The difference of the frequencies in $K$ and $\bar{K}$ will be discussed in Sec.~\ref{sec:time-evol-major_ferro}.

For comparison, we compute the time evolution for the quench from the forced-FM state.
Figures~\ref{fig_waveletF}(d), \ref{fig_waveletF}(e), and~\ref{fig_waveletF}(f) display the time evolutions of $M^z$, $K$, and $\bar{K}$, respectively, at $h/|J|=0.15$ well above $h_c$.
All the results show damped oscillations, where the amplitude decreases and the frequency is almost constant.
The damping occurs because the total $S^z$ is not a good quantum number in the Kitaev model.
The important point is that the time evolutions in $K$ and $\bar{K}$ are identical except for the sign, indicating that the dynamics of the Majorana fermions $\{a, b\}$ and $\{\bar{a}, \bar{b}\}$ is indistinguishable.
The same behavior is also observed for the quench from the forced-FM state in the AFM case (not shown)
~\footnote{
For the field quench from the forced-FM phase in the AFM Kitaev model, we find that the numerical precision becomes worse compared to other cases, as shown in Appendix~\ref{sec:conservation-law}.}.
These imply that the spin fractionalization is not observed in the field quench from the forced-FM state, where there is no fractionalization in the equilibrium state, as shown in Sec.~\ref{sec:result_h=0}.
The transient dynamics is understood simply by the spin precession.
This is in sharp contrast to the fractional dynamics observed in the quench from the QSL in Figs.~\ref{fig_waveletF}(b) and~\ref{fig_waveletF}(c).

\subsubsection{Antiferromagnetic case: Kitaev QSL}\label{sec:antif-case:-kita}

\begin{figure*}[t]
 \begin{center}
  \includegraphics[width=2\columnwidth,clip]{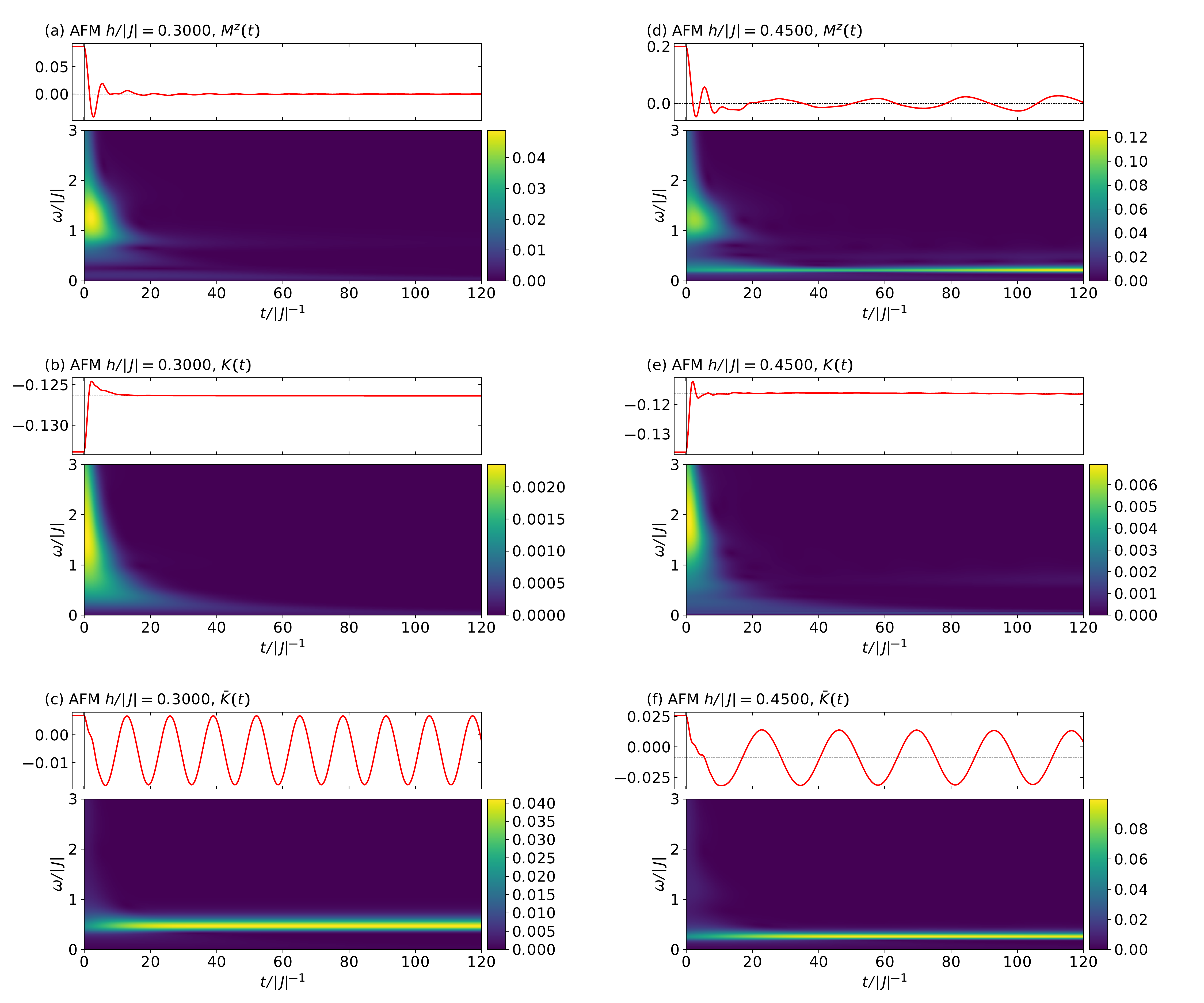}
  \caption{
Time evolutions of (a) $M^z$, (b) $K$, and (c) $\bar{K}$, and their wavelet scalograms in the AFM Kitaev model for the field quench from $h/J=0.3$ (Kitaev QSL state).
(d)--(f) Corresponding plots from $h/J=0.45$ (intermediate QSL state).
The dotted lines represent the long-time averages of each quantity.
}
  \label{fig_waveletAF1}
 \end{center}
\end{figure*}

Next, we study the time evolution in the AFM Kitaev model, which exhibits two different QSLs before entering the forced-FM phase in the static magnetic field, as described in Sec.~\ref{sec:result_h=0}.
In this section, we focus on the results for the field quench from the low-field Kitaev QSL state below $h_{c1}/|J|\simeq 0.417$.

Figures~\ref{fig_waveletAF1}(a)--\ref{fig_waveletAF1}(c) show the time evolutions in the field quench from $h/|J|=0.3$.
As shown in Fig.~\ref{fig_waveletAF1}(a), the magnetization damps rapidly for $t/|J|^{-1}\lesssim 10$.
The early-stage dynamics has a board spectrum for $1\lesssim \omega/|J| \lesssim 3$ as observed in the wavelet scalogram.
Similar behavior is found in the time evolution of $K$ shown in Fig.~\ref{fig_waveletAF1}(b).
This result implies that the dynamics of the magnetization is predominantly ascribed to the excitation of the Majorana fermions $\{a, b\}$.
On the other hand, $\bar{K}$ shows a long-lived quasi-coherent oscillation, as shown in Fig.~\ref{fig_waveletAF1}(c).
These results suggest that the dynamics of the Majorana fermions $\{a, b\}$ and $\{\bar{a}, \bar{b}\}$ are well separated and exhibit distinct characteristics; the former emerges as higher-energy excitations with a short lifetime, whereas the latter as low-energy but long-lived excitations.

The distinct behavior is qualitatively understood in the original spin picture.
As shown in Eq.~\eqref{eq:7}, $K$ represents the NN correlation for the interacting spin component, whereas $\bar{K}$ corresponds to that for the noninteracting one.
Before quenching, the magnetic field renders spins aligned against the AFM interactions.
After the forced alignment is released by the quench, the energy accumulated in the interacting spin component is transferred to the noninteracting ones, which can fluctuate more freely.
Such behavior is indeed seen in the time evolutions in Figs.~\ref{fig_waveletAF1}(b) and \ref{fig_waveletAF1}(c).
This would be the reason why the long-lived oscillation appears in $\bar{K}$, whereas it is absent in $K$.
Similar behavior is found also in the FM case although the accumulation energy and its transfer are much smaller than the AFM case; see Figs.~\ref{fig_waveletF}(b) and~\ref{fig_waveletF}(c).
Thus, the energy transfer between the two kinds of Majorana fermions in the early-time stage and the resultant long-lived oscillation in $\bar{K}$ are common features to the QSLs driven by the energy transfer between the itinerant and localized Majorana fermions.

On the other hand, qualitative difference is also present between the FM and AFM cases in the time evolution of $M^z$;
namely, it shows a long-lived oscillation in the FM case [Fig.~\ref{fig_waveletF}(a)], but damps quickly in the AFM case [Fig.~\ref{fig_waveletAF1}(a)].
This is understood from the difference in the dynamical spin structure factor ${\cal S}(\bm{q}=0,\omega)$ in the equilibrium state at $h=0$~\cite{PhysRevLett.112.207203,PhysRevB.92.115127} discussed in Sec.~\ref{sec:model} (see also Fig.~\ref{fig_sw} in Appendix~\ref{sec:valid-time-depend}).
As mentioned in Sec.~\ref{sec:time-evol-magn_ferro}, the low-energy peak in ${\cal S}(\bm{q}=0,\omega)$ leads to the slow oscillation of $M^z$ in the FM case.
On the other hand, such a low-energy coherent peak is absent in the AFM Kitaev model, leaving only a high-energy broad structure, as described in Sec.~\ref{sec:model}.
This results in the absence of the low-energy component observed in Fig.~\ref{fig_waveletAF1}(a).

\subsubsection{Antiferromagnetic case: intermediate QSL}

More conspicuous time evolutions are found in the case of the quench from the intermediate QSL state between $h_{c1}/|J|\simeq0.417$ and $h_{c2}/|J|\simeq0.503$.
The representative results are shown for $h/|J|=0.45$ in Figs.~\ref{fig_waveletAF1}(d)--\ref{fig_waveletAF1}(f).
As shown in the upper panel of Fig.~\ref{fig_waveletAF1}(d), the magnetization exhibits two-step transient dynamics: a high-frequency oscillation in the early-time stage and a slower oscillation in the longer-time range.
This peculiar time evolution is clearly observed in the wavelet scalogram.
The high-energy broad structure appears for $1\lesssim\omega/|J|\lesssim3$, which quickly decays for $t/|J|^{-1}\lesssim 10$.
This can be attributed to the Majorana fermions $\{a, b\}$ because a similar scalogram is observed for $K$ in Fig.~\ref{fig_waveletAF1}(e).
Similar correspondence was observed also for the low-field QSL case in Figs.~\ref{fig_waveletAF1}(a) and \ref{fig_waveletAF1}(b).
On the other hand, the scalogram in Fig.~\ref{fig_waveletAF1}(d) indicates the enhancement of the low-energy component with a narrow peak at $\omega/|J|\sim 0.2$ caused by the elapse of time.
Similar behavior is observed in $\bar{K}$ shown in Fig.~\ref{fig_waveletAF1}(f), suggesting that this is ascribed to the Majorana fermions $\{\bar{a}, \bar{b}\}$.
Thus, the dynamics of the magnetization reflects both features of $\{a,b\}$ and $\{\bar{a},\bar{b}\}$ in this case.
This behavior will be discussed in Sec.~\ref{sec:antif-case:-interm}.

\subsection{Time evolution of Majorana bands}\label{sec:time-evol-major}

In the previous section, we clarified the time evolutions of the magnetization and spin correlations, and elucidated how the spin fractionalization manifests itself in the transient dynamics through the distinct lifetimes of the quasiparticles.
The quasiparticles, Majorana fermions $\{a, b\}$ and $\{\bar{a}, \bar{b}\}$, are hybridized in the presence of the magnetic field, and the hybridization is retained even after the field quench.
This induces the interesting dynamics in the cases of quenching from the QSLs with fractional excitations.
In this section, we analyze directly the time evolution of the Majorana fermion states to further clarify the characteristic nonequilibrium dynamics.

In the following, we calculate the time dependence of the Majorana band structure.
To define the Majorana band, we introduce complex fermions to diagonalize the Majorana MF Hamiltonian as
\begin{align}
{\cal H}^{\rm MF}=\sum_{\bm{k}}\sum_{\mu=-1,-2}E_{\bm{k}\mu}\left(f_{\bm{k}\mu}^\dagger f_{\bm{k}\mu}-\frac{1}{2}\right),
\label{eq:H_fermion}
\end{align}
where $f_{\bm{k}\mu}^\dagger$ and $f_{\bm{k}\mu}$ are creation and annihilation operators of the complex fermions, respectively, and $\mu$ is the band index.
The details are given in Supplemental Material in Ref.~\cite{Nasu2018mag}.
In the equilibrium state, the eigenenergy $E_{\bm{k}\mu}=2\varepsilon_{\bm{k}\mu}$ is negative for the two bands with $\mu=-1$ and $-2$ [$\varepsilon_{\bm{k}\mu}$ is the eigenvalue of Eq.~\eqref{eq:Schrodinger_eq}].
The ground state is described by the fully occupied state of the two bands, whose energy is given by
\begin{align}
 E_g=\sum_{\bm{k}}\sum_{\mu=-1,-2}\frac{E_{\bm{k}\mu}}{2},
\end{align}
and the excited states are expressed by annihilation of the $f$ fermions.
The density of states (DOS) of the occupied bands is defined as
\begin{align}
 D_{\rm occ}(\omega)=\frac{1}{N}\sum_{\bm{k}}\sum_{\mu=-1,-2}\delta(\omega-E_{\bm{k}\mu}).
\label{eq:DOS}
\end{align}

In the time-dependent Majorana MF method, the filled bands evolve adiabatically, and therefore, the two bands remain to be occupied in the time development of the ground state after the field quench.
To track the hybridization of the bands in the elapse of time, we introduce the occupations for two kinds of Majorana fermions $\{a,b\}$ and $\{\bar{a}, \bar{b}\}$ as
\begin{align}
 n_{\bm{k}\mu}=\bras{\phi_{\bm{k}\mu}}\left(a_{\bm{k}}^\dagger a_{\bm{k}} +b_{\bm{k}}^\dagger b_{\bm{k}}\right)\kets{\phi_{\bm{k}\mu}},\label{eq:8}
\end{align}
and
\begin{align}
 \bar{n}_{\bm{k}\mu}=\bras{\phi_{\bm{k}\mu}}\left(\bar{a}_{\bm{k}}^\dagger \bar{a}_{\bm{k}} + \bar{b}_{\bm{k}}^\dagger \bar{b}_{\bm{k}}\right)\kets{\phi_{\bm{k}\mu}},\label{eq:10}
\end{align}
 respectively, where $c_{\bm{k}}$ ($c=a,b,\bar{a},\bar{b}$) is the Fourier transform of $c_j$.
Note that $\bar{n}_{\bm{k}\mu}=1-n_{\bm{k}\mu}$.
In the presence of the hybridization,
$n_{\bm{k}\mu}$ and $\bar{n}_{\bm{k}\mu}$ take values between 0 and 1.
Therefore, as a measure of the hybridization, we compute the time evolution of
\begin{align}
 w_{\bm{k}\mu}=n_{\bm{k}\mu}\bar{n}_{\bm{k}\mu}.
\label{eq:9}
\end{align}
This quantity does not depend on the band, and hence, we drop the band index $\mu$ and denote it as $w_{\bm{k}}$ hereafter.

\subsubsection{Initial state at $t=0$}
\label{sec:band_t=0}

\begin{figure}[t]
 \begin{center}
  \includegraphics[width=\columnwidth,clip]{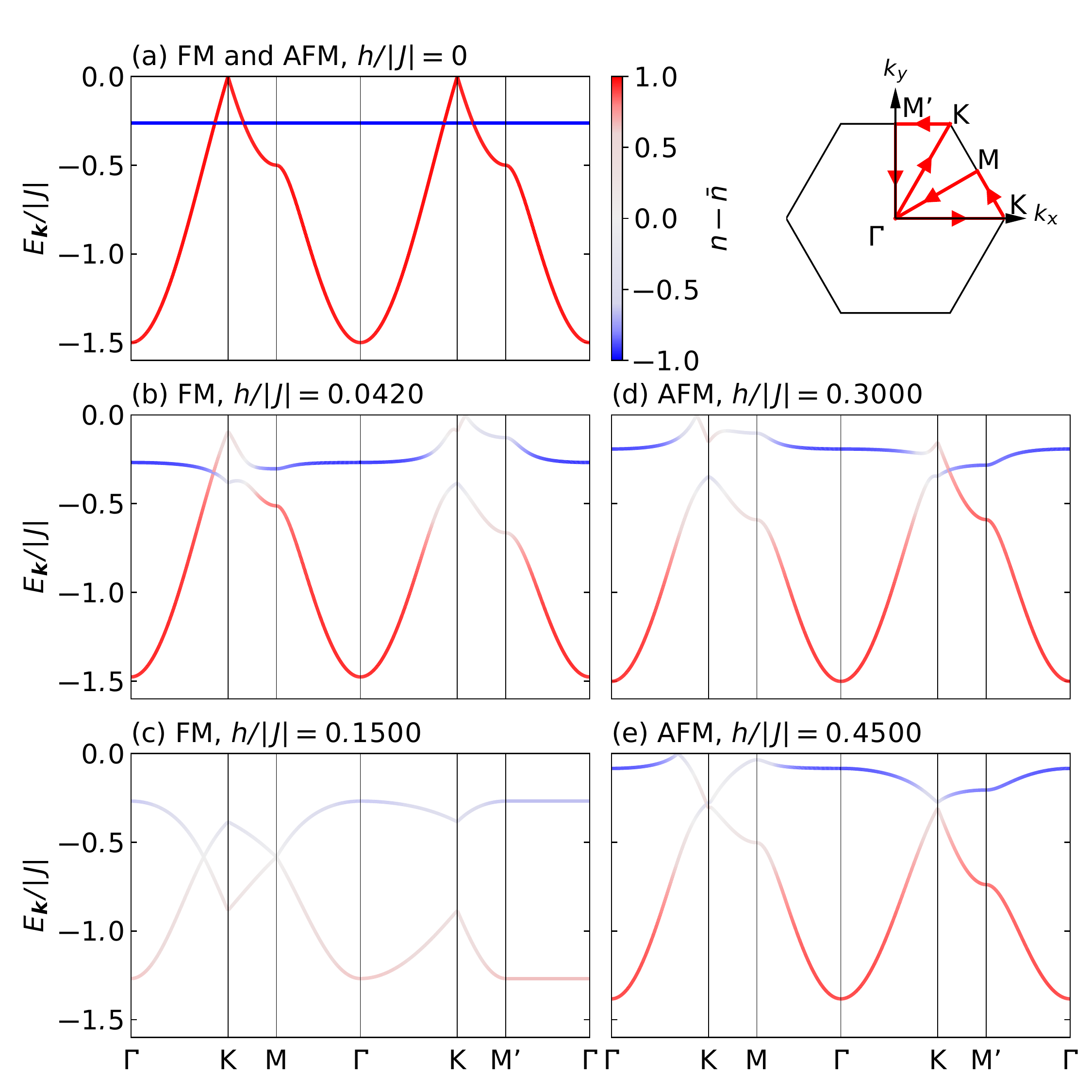}
  \caption{
 Majorana band structures in the static magnetic field for (a) $h/|J|=0$ (Kitaev QSL) in both FM and AFM cases, (b) $h/|J|=0.0420$ (Kitaev QSL) in the FM case, (c) $h/|J|=0.15$ (forced FM) in the FM case, (d) $h/|J|=0.3$ (Kitaev QSL) in the AFM case, and (e) $h/|J|=0.45$ (intermediate QSL) in the AFM case.
The dispersions are shown along the red lines in the Brillouin zone depicted in the top right.
The color stands for $n_{\bm{k}\mu} -\bar{n}_{\bm{k}\mu}$ for the corresponding band $\mu$ with the wavenumber $\bm{k}$.
}
  \label{fig_disp_hzero}
 \end{center}
\end{figure}

Before showing the results of the time-dependent Majorana bands, we briefly discuss the Majorana bands in the equilibrium states under the magnetic field.
Figure~\ref{fig_disp_hzero} shows the Majorana band structures at several magnetic fields in the FM and AFM Kitaev models, obtained by the static Majorana MF theory.
At $h=0$, where the Majorana MF theory gives the exact solution, there are dispersive and flat bands, which originate from the itinerant Majorana fermions $\{a,b\}$ and the localized ones $\{\bar{a}, \bar{b}\}$, respectively [Fig.~\ref{fig_disp_hzero}(a)].
The dispersive band has nodal points with linear dispersions at the K points.
In this case, the dispersions in the FM and AFM cases are identical.
By introduction of $h$, these two kinds of bands are hybridized, and exhibit anticrossing behavior, as shown Figs.~\ref{fig_disp_hzero}(b), \ref{fig_disp_hzero}(d), and \ref{fig_disp_hzero}(e);
the flat band becomes dispersive with a narrow bandwidth, whose typical energy is pushed up, while the anticrossed dispersive band is pushed down, in a different manner between the FM and AFM cases.
At the same time, the nodal points shift horizontally in the Brillouin zone to the M' ($\Gamma$) point in the FM (AFM) case.
In the AFM case, as discussed in the previous study~\cite{Nasu2018mag}, the phase transition from the low-field QSL to the intermediate QSL is accompanied by a topological change of the Majorana bands.
In both FM and AFM cases, the nodal points disappear and the Majorana bands are fully gapped in the forced-FM phase, as shown in Fig.~\ref{fig_disp_hzero}(c) for the FM case.

\subsubsection{Ferromagnetic case}
\label{sec:time-evol-major_ferro}

\begin{figure}[t]
 \begin{center}
  \includegraphics[width=\columnwidth,clip]{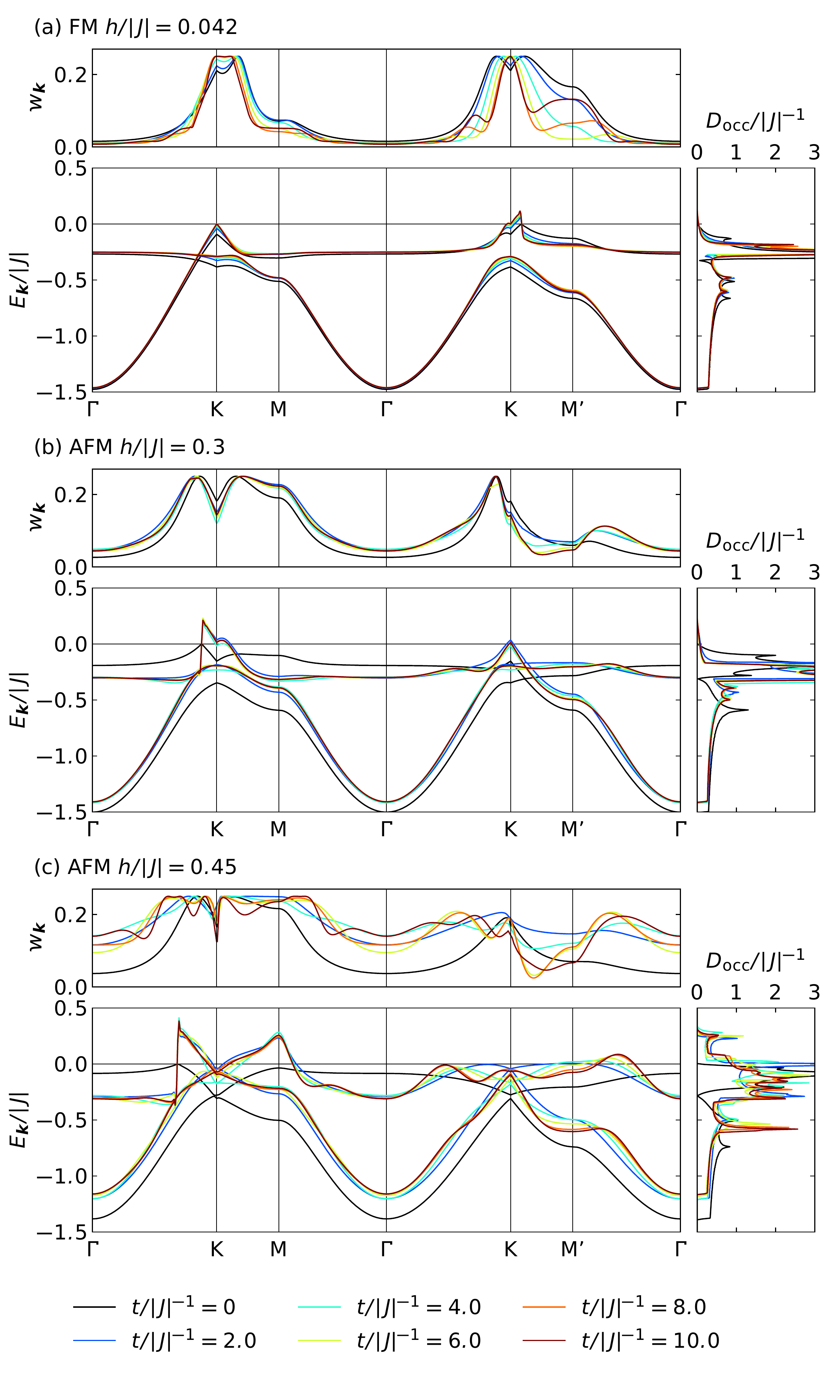}
  \caption{
Time evolutions of the dispersion $E_{\bm{k}\mu}$ [Eq.~\eqref{eq:H_fermion}], the hybridization between two kinds of Majorana fermions $w_{\bm{k}}$ [Eq.~\eqref{eq:9}], and the occupied DOS $D_{\rm occ}$ [Eq.~\eqref{eq:DOS}] in the early-time stage after the field quench from (a)  $h/|J|=0.0420$ (Kitaev QSL) in the FM case, (b) $h/|J|=0.3$ (Kitaev QSL) in the AFM case, and (c) $h/|J|=0.45$ (intermediate QSL) in the AFM case.
The dispersions are shown along the red line in the Brillouin zone depicted in Fig.~\ref{fig_disp_hzero}.
}
  \label{fig_disp_dos}
 \end{center}
\end{figure}

\begin{figure*}[t]
 \begin{center}
  \includegraphics[width=2\columnwidth,clip]{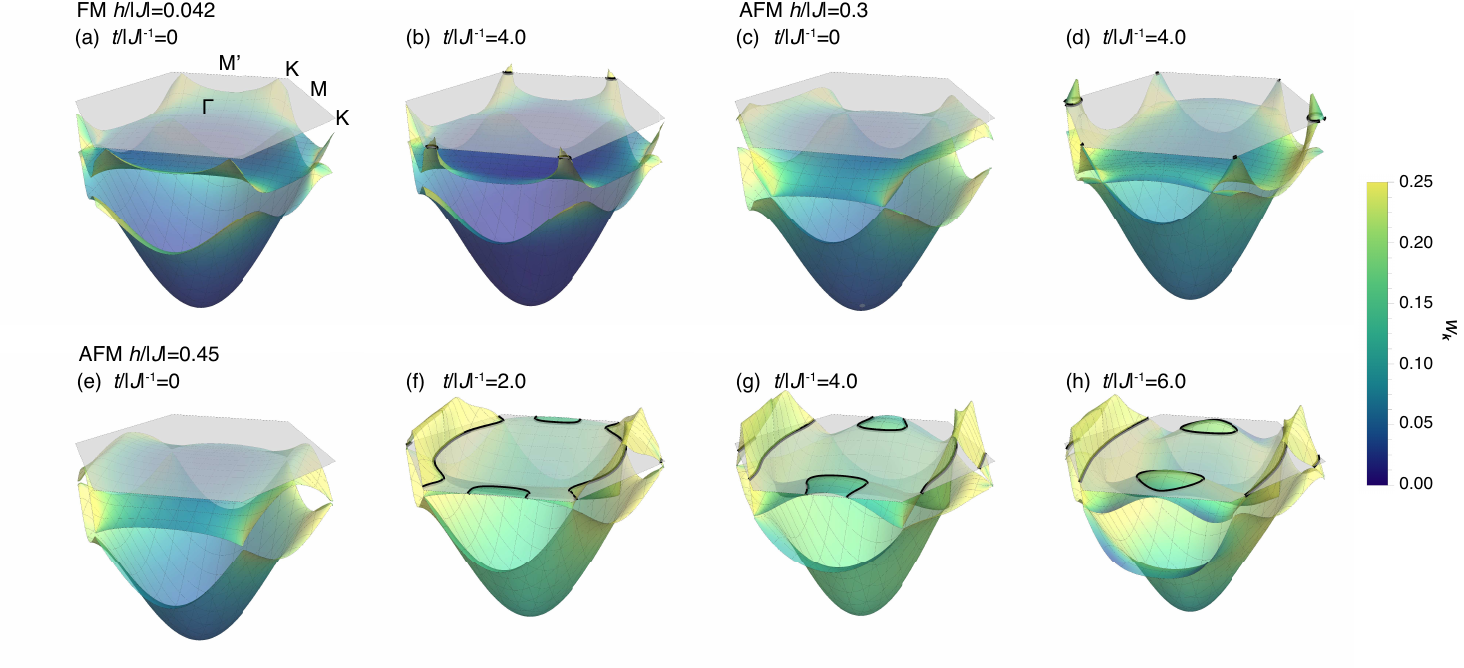}
  \caption{
Three-dimensional plots of the time evolution of the dispersion $E_{\bm{k}\mu}$ [Eq.~\eqref{eq:H_fermion}] in the first Brillouin zone for the field quench from (a),(b) $h/|J|=0.0420$ (Kitaev QSL) in the FM case, (c),(d) $h/|J|=0.3$ (Kitaev QSL) in the AFM case, and (e)--(h) $h/|J|=0.45$ (intermediate QSL) in the AFM case.
The color represents the hybridization between two kinds of Majorana fermions $w_{\bm{k}}$ [Eq.~\eqref{eq:9}] (see also Fig.~\ref{fig_disp_dos}).
The gray hexagons represent the first Brillouin zone, and the black curves on them indicate the zero-energy level corresponding to the ``Fermi surfaces'' (see also Fig.~\ref{fig_fermis}).
}
  \label{fig_band}
 \end{center}
\end{figure*}

\begin{figure}[t]
 \begin{center}
  \includegraphics[width=\columnwidth,clip]{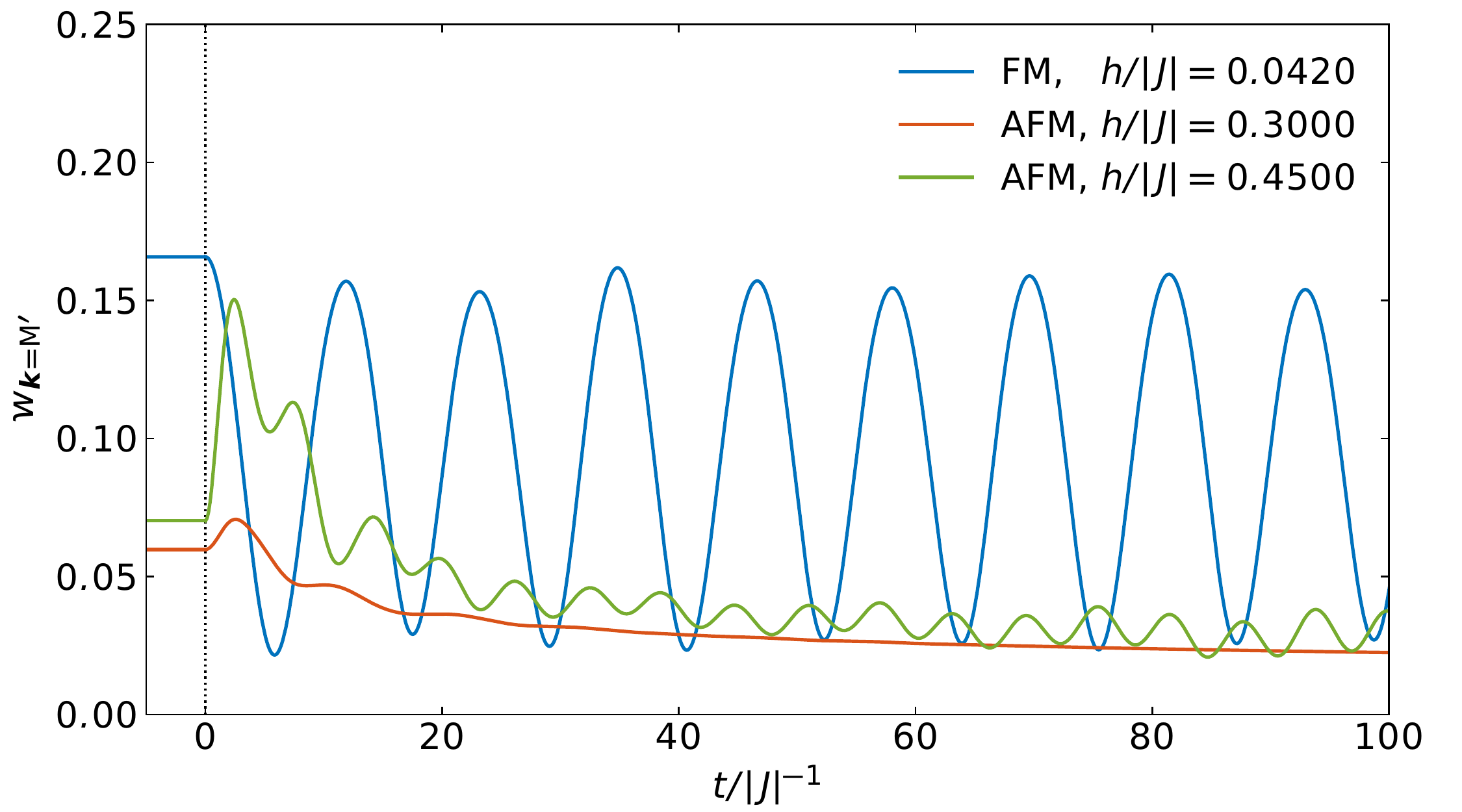}
  \caption{
Time evolutions of the hybridization between the two kinds of Majorana fermions, $w_{\bm{k}}$ in Eq.~\eqref{eq:9}, at the M' point for three cases in Fig.~\ref{fig_disp_dos}.
}
  \label{fig_w_t_dep}
 \end{center}
\end{figure}

First, we discuss the time evolution of the Majorana fermion states in the FM case.
Figure \ref{fig_disp_dos}(a) shows the time evolution of the band structure after the field quench from the Kitaev QSL state at $h/|J|=0.0420$, which is just below $h_c/|J|\simeq 0.0421$ [see Fig.~\ref{fig_hdep_full}(a)].
The overall band structures in the first Brillouin zone are shown for $t/|J|^{-1} = 0$ and $4.0$ in Figs.~\ref{fig_band}(a) and \ref{fig_band}(b), respectively.

Before quenching ($t=0$), as discussed in Sec.~\ref{sec:band_t=0}, the anticrossing behavior is observed between the dispersive and flat bands, which are dominated by $\{a,b\}$ and $\{\bar{a},\bar{b}\}$, respectively.
In this state, the hybridization between the two kinds of Majorana fermions, $w_{\bm{k}}$, takes a large value around the anticrossing regions near the K point, as plotted in the top panel of Fig.~\ref{fig_disp_dos}(a) [see also the color plot in Fig.~\ref{fig_disp_hzero}(b)].

After quenching, the band structure shows substantial time evolution around the anticrossing regions, as shown in Fig.~\ref{fig_disp_dos}(a).
In the same regions, two kinds of Majorana fermions $\{a,b\}$ and $\{\bar{a},\bar{b}\}$ remain to be strongly hybridized in the elapse of time, as shown in the top panel of Fig.~\ref{fig_disp_dos}(a).
We note that the hybridization $w_{\bm{k}}$ shows rather large time dependence around the M' point.
We show the time evolution of $w_{\bm{k}}$ at the M' point in Fig.~\ref{fig_w_t_dep}.
The result indicates that $w_{\bm{k}}$ shows a quasi-coherent oscillation for a longer time. This corresponds to the long-lived oscillation of the magnetization observed in Fig.~\ref{fig_waveletF}(a), because the magnetization is described by the hybridization of the two Majorana fermions as $M^z=\frac{i}{2}\left(\sum_{j\in A}\means{a_j\bar{a}_j}-\sum_{j'\in B}\means{b_{j'}\bar{b}_{j'}}\right)$.
The hybridization and anticrossing also account for the difference of the frequencies of long-lived oscillations in $K$ and $\bar{K}$ discussed in Sec.~\ref{sec:time-evol-magn_ferro} as follows.
As shown in Fig.~\ref{fig_disp_hzero}(b), the band dominated by $\{a,b\}$ ($\{\bar{a},\bar{b}\}$) is pushed down (up) at $t=0$; we confirm that this feature is roughly retained in the time evolution after quenching (not shown).
The result suggests that the dominant energy scale for $K$ by $\{a,b\}$ is higher than that for $\bar{K}$ by $\{\bar{a},\bar{b}\}$, which is consistent with the results in Figs.~\ref{fig_waveletF}(b) and \ref{fig_waveletF}(c).

\begin{figure}[t]
 \begin{center}
  \includegraphics[width=\columnwidth,clip]{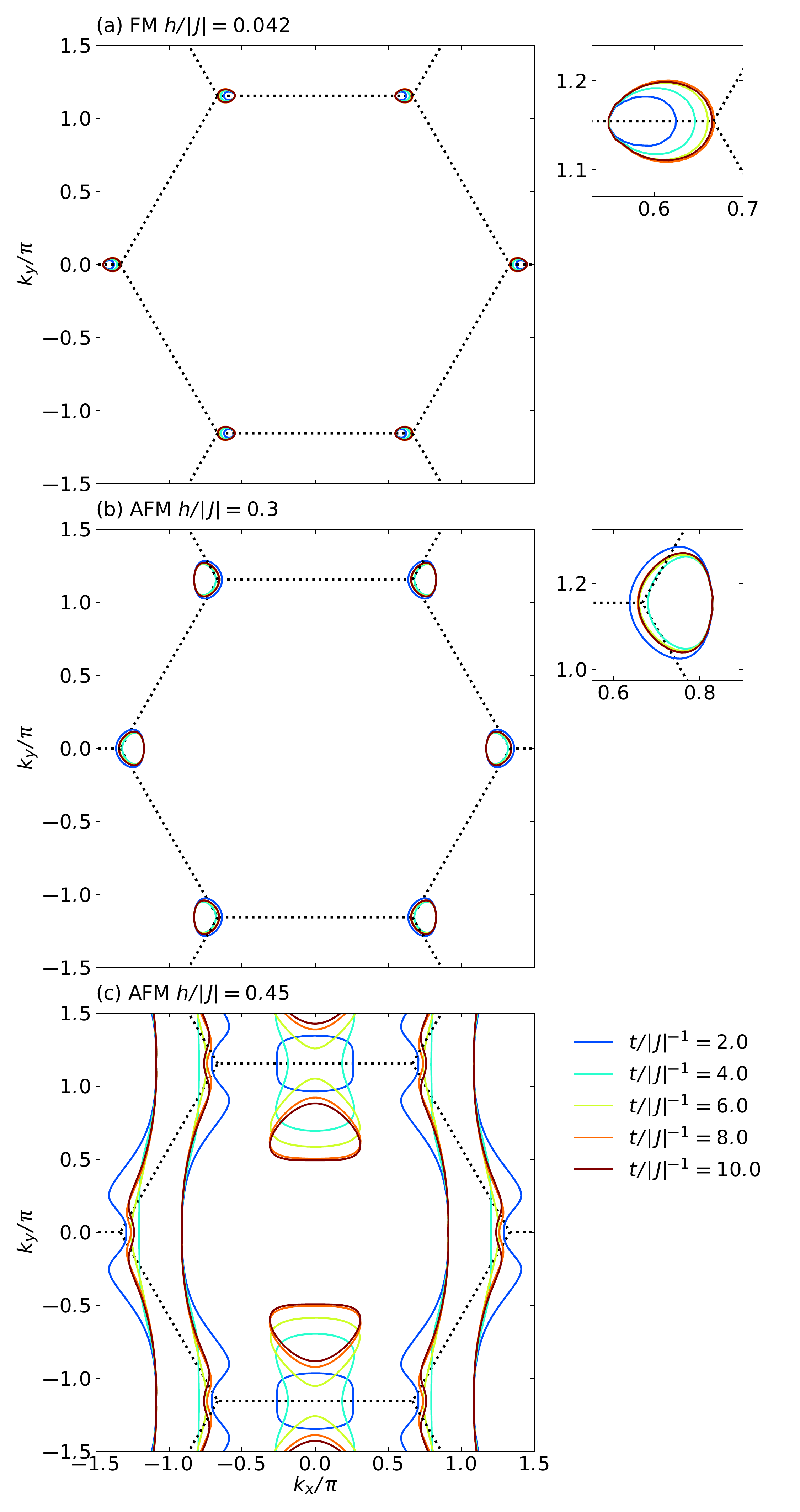}
  \caption{
Time evolution of the ``Fermi surfaces'' in the early-time stage after the field quench from (a) $h/|J|=0.0420$ (Kitaev QSL) in the FM case, (b) $h/|J|=0.3$ (Kitaev QSL) in the AFM case, and (c) $h/|J|=0.45$ (intermediate QSL) in the AFM case.
The dotted lines indicate the Brillouin zone boundaries.
Extended plots of the Fermi surface around the K point are shown in the right panels of (a) and (b).
}
  \label{fig_fermis}
 \end{center}
\end{figure}

The most interesting point in the time evolution of the Majorana states is that the occupied band is pushed above zero energy around the K point after the quench, as shown in Fig.~\ref{fig_disp_dos}(a).
As quasiparticles at the zero-energy level can annihilate without energy cost, the zero energy is regarded as the Fermi level in this system.
Thus, the appearance of the positive-energy band indicates the appearance of the transient ``Fermi surface''.
To show this more clearly, we present the time evolution of the Fermi surface in Fig.~\ref{fig_fermis}(a).
The point node on the K-M' line at $t=0$ develops into an oval-shaped Fermi surface after the field quench [see also Figs.~\ref{fig_band}(a) and \ref{fig_band}(b)].
The Fermi surface grows while time elapses, as shown in the extended plot around the K point in the right panel of Fig.~\ref{fig_fermis}(a).

We also calculate the time evolution of the DOS $D_{\rm occ}(\varepsilon)$ in Eq.~\eqref{eq:DOS}.
The result is shown in the right panel of Fig.~\ref{fig_disp_dos}(a).
As expected from the appearance of the transient Fermi surface, we find that $D_{\rm occ}(\varepsilon=0)$ becomes nonzero for $t>0$.
Since the low-energy excitations at the Fermi level play an essential role in the thermodynamics and transport phenomena at low temperatures, the appearance of the transient Fermi surface by the field quench will affect the low-energy dynamical properties significantly (see Sec.~\ref{sec:discussion}).

It is worth noting that our Fermi surface does not define the boundary between the occupied and unoccupied states since the bands shown here remain to be occupied in the adiabatic time evolution as mentioned above.
When we take into account the energy dissipation, the transient occupied states above the Fermi surfaces will become unoccupied in longer time scales.
Nonetheless, we expect that the transient Fermi surfaces lead to interesting behaviors before dissipation, as discussed in Sec.~\ref{sec:discussion}.

\subsubsection{Antiferromagnetic case: Kitaev QSL}

Next, we present the time evolution of the Majorana fermion states for the AFM Kitaev model.
Figure~\ref{fig_disp_dos}(b) shows the Majorana band structure, the hybridization $w_{\bm{k}}$, and the DOS $D_{\rm occ}$ for the field quench from the Kitaev QSL state at $h/|J|=0.3$ below $h_{c1}/|J|\simeq0.417$.
See also Figs.~\ref{fig_band}(c) and \ref{fig_band}(d) for the three-dimensional plots of the Majorana band dispersions.
As shown in Fig.~\ref{fig_disp_dos}(b), the overall dispersions are shifted to a high-energy side immediately after quenching the magnetic field.
This can be understood from the energy redistribution between the quasiparticles $\{a,b\}$ and $\{\bar{a},\bar{b}\}$ discussed in Sec.~\ref{sec:antif-case:-kita} as follows.
After the quench, the energy of interacting spin components corresponding to $K$ is transferred to that of noninteracting components corresponding to $\bar{K}$.
As the dispersion with the wide bandwidth is predominantly composed of the Majorana fermions $\{a,b\}$, the upward shift can be attributed to such a sudden decrease of $|K|$.
We note that a similar effect is present also in the FM case, but it is much smaller and difficult to see in Fig.~\ref{fig_disp_dos}(a).

After the sudden change, the band structure shows relatively large time evolution in the region where the hybridization $w_{\bm{k}}$ is large [see the top panel of Fig.~\ref{fig_disp_dos}(b)], similar to the FM case in Fig.~\ref{fig_disp_dos}(a).
We plot the time evolution of $w_{\bm{k}}$ at the M' point in Fig.~\ref{fig_w_t_dep}.
In contrast to the FM case, this quantity shows a rapid increase in the early-time stage followed by a decay with small oscillations.
The result may correspond to the absence of the low-frequency oscillation in the magnetization in Fig.~\ref{fig_waveletAF1}(a), since the magnetization is related with the hybridization as discussed in Sec.~\ref{sec:time-evol-major_ferro}.

Furthermore, similar to the FM case, we find the appearance of the positive-energy band and the transient Fermi surface around the K point [Figs.~\ref{fig_disp_dos}(b), \ref{fig_band}(d), and \ref{fig_fermis}(b)].
This is a common feature of the quench from the Kitaev QSL state in the FM and AFM cases.
Meanwhile, the Fermi surface in the AFM Kitaev model is larger than that in the FM one.
This is again understood as the redistribution of the exchange energy in the interacting spin components accumulated in the magnetic field; the accumulation energy in AFM case should be larger than that in the FM one, which leads to the larger size of the transient Fermi surface.

\subsubsection{Antiferromagnetic case: intermediate QSL}\label{sec:antif-case:-interm}

Figures~\ref{fig_disp_dos}(c) and \ref{fig_band}(e)--\ref{fig_band}(h) show the time evolution of the band dispersions for the quench from the intermediate QSL phase in the AFM Kitaev model at $h/|J|=0.45$.
In contrast to the previous cases for the Kitaev QSLs, the band structure is largely  reconstructed in a wide energy range after a sudden upshift.
Accordingly, the DOS also shows strong time dependence, as plotted in the right panel of Fig.~\ref{fig_disp_dos}(c).
We note that, in this case, the hybridization $w_{\bm{k}}$ becomes relatively large in the whole Brillouin zone after the quench, including the region away from the anticrossings.
The large hybridization allows us to observe two kinds of Majorana dynamics via the magnetization, as seen in Fig.~\ref{fig_waveletAF1}(d).

Corresponding to the drastic time evolution, there appears the positive-energy band in many portions of the Brillouin zone, such as around the M point and along the $\Gamma$-K and $\Gamma$-M' lines, as shown in Fig.~\ref{fig_disp_dos}(c).
This leads to the transient Fermi surfaces, whose time dependence is more conspicuous compared to the previous cases, as shown in Fig.~\ref{fig_fermis}(c) [see also Figs.~\ref{fig_band}(e)--\ref{fig_band}(h)].
After quenching, a large Fermi surface appears immediately along the $k_y$ direction centered around the M point, and a small pocket also appears around the M' point.
The latter is an open Fermi surface on the Brillouin zone boundary, but changes its topology into a closed one at $t/|J|^{-1} \sim 5$.
This is regarded as a dynamical version of the ``Lifshitz transition'' of the Majorana fermion system.

\subsection{Time evolution of Majorana density of states}\label{sec:time-evol-majorDOS}
\begin{figure}[t]
 \begin{center}
  \includegraphics[width=\columnwidth,clip]{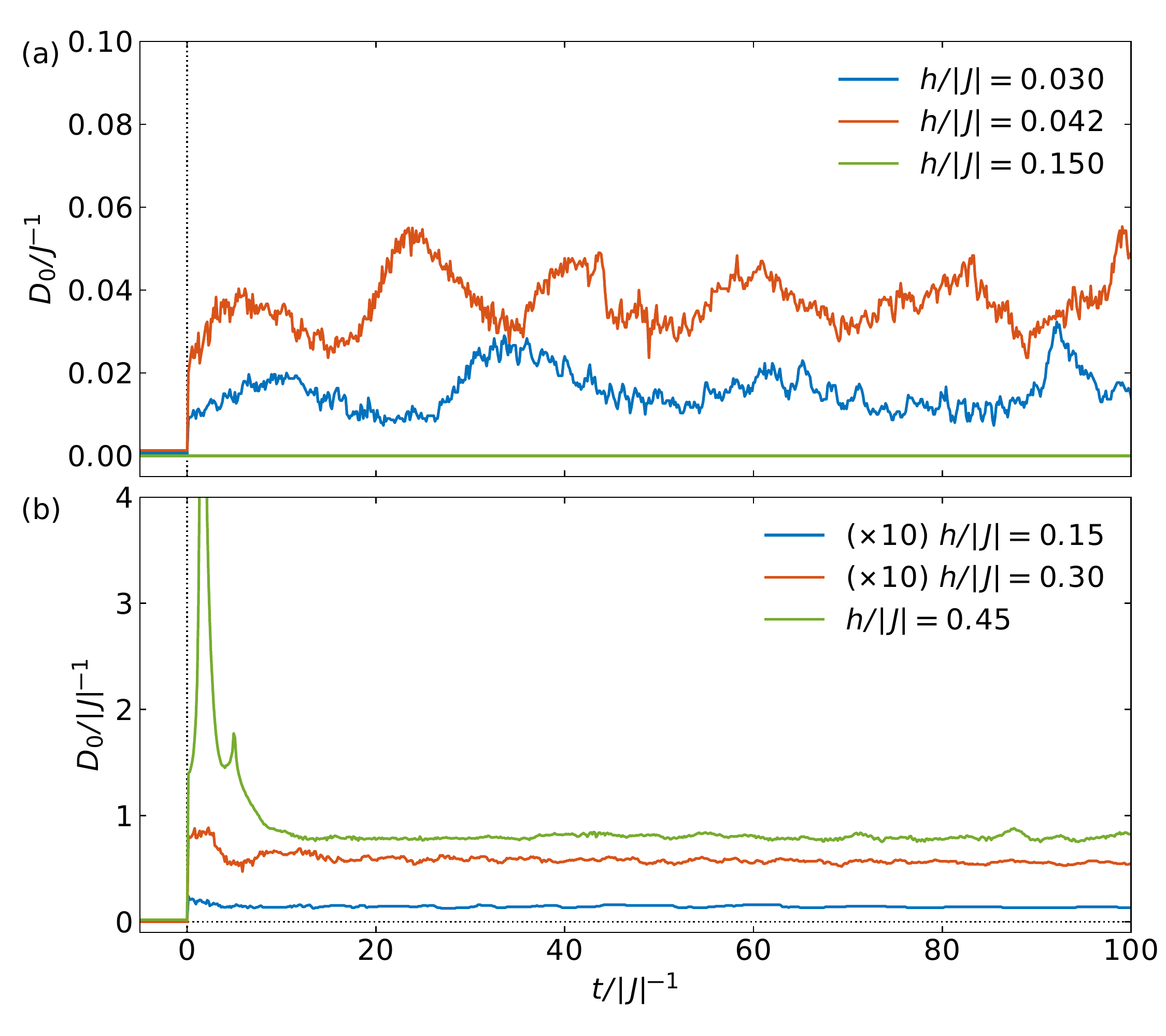}
  \caption{
Time evolution of the DOS at the Fermi level in (a) the FM Kitaev model and (b) the AFM one.
}
  \label{fig_fermi_dos}
 \end{center}
\end{figure}

\begin{figure}[t]
 \begin{center}
  \includegraphics[width=\columnwidth,clip]{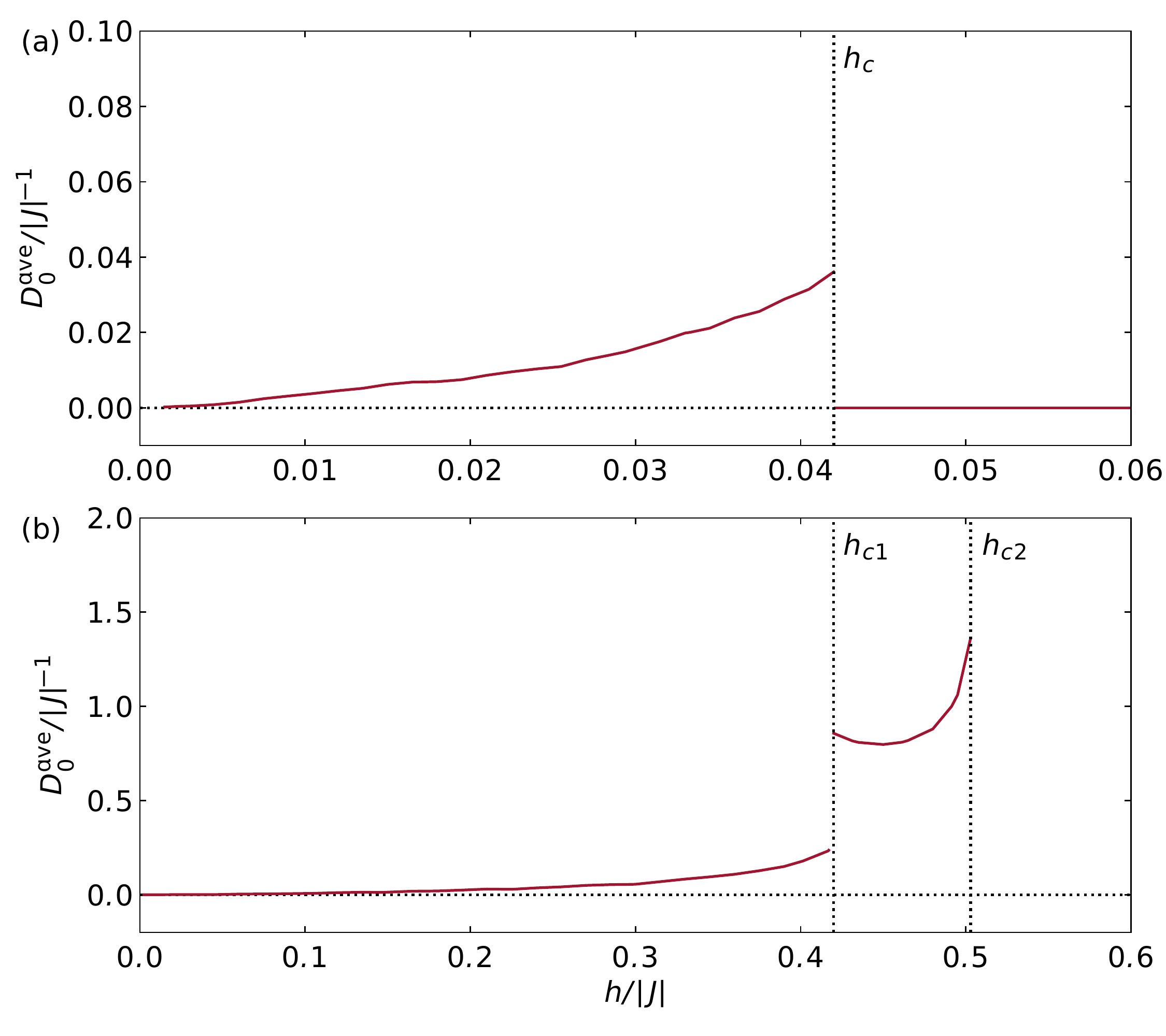}
  \caption{
Initial field dependence of the long-time average of the DOS at the Fermi level, $D_0^{\rm ave}$, in the time range of $50<t/|J|^{-1}<200$ for (a) the FM Kitaev model and (b) the AFM one.
}
  \label{fig_h_dep_dos}
 \end{center}
\end{figure}

In the previous section, we found the appearance of the transient Fermi surfaces by the magnetic-field quench.
The corresponding time evolutions of the DOS were shown in the right panels of Fig.~\ref{fig_disp_dos}.
In this section, we discuss the longer-time dynamics of the DOS.
In Fig.~\ref{fig_fermi_dos}, we show the time evolution of the DOS at the Fermi level, $D_0 = D_{\rm occ}(\omega=0)$, up to $t/|J|^{-1}=100$.
In the FM case [Fig.~\ref{fig_fermi_dos}(a)], before quenching, $D_0$ is zero for all $h$ because the system is ``semimetal'' with the point nodes in the Kitaev QSL phase for $h<h_c$ and it is gapped in the forced-FM phase for $h>h_c$ ($h_c/|J|\simeq0.0421$) [see Figs.~\ref{fig_disp_hzero}(a)--\ref{fig_disp_hzero}(c)].
After quenching, $D_0$ suddenly becomes nonzero by the quench from the initial field $h<h_c$.
In the wider time range, $D_0$ stays almost constant with a slow fluctuation, as plotted in Fig.~\ref{fig_fermi_dos}(a).
The overall value of $D_0$ increases with an increase of $h$, as shown in Fig.~\ref{fig_fermi_dos}(a).
On the other hand, in the case of the field quench from $h>h_c$, $D_0$ remains zero, as exemplified at $h/|J|=0.15$ in Fig.~\ref{fig_fermi_dos}(a).
This is due to the presence of the gap persisting in the time evolution from the forced-FM state.
The contrasting results are explicitly shown by plotting the long-time average of the DOS, $D_0^{\rm ave}$, as a function of the initial field $h$ in Fig.~\ref{fig_h_dep_dos}(a); here we compute the average in the time range of $50<t/|J|^{-1}<200$.
As shown in Fig.~\ref{fig_h_dep_dos}(a), $D_0^{\rm ave}$ becomes nonzero for $h>0$ and monotonically increases while increasing $h$, but it vanishes above $h_c$.

Figure~\ref{fig_fermi_dos}(b) shows the time dependence of $D_0$ in the AFM case.
After the field quench, large changes are observed compared to the FM case, particularly in the case of the quench from the intermediate QSL state between $h_{c1}/|J|\simeq0.417$ and $h_{c2}/|J|\simeq0.503$, as discussed in the previous section.
While $D_0$ is strongly enhanced in the early-time stage, it quickly converges to almost constant for the longer time $t/|J|^{-1}\gtrsim 10$.
Figure~\ref{fig_h_dep_dos}(b) shows the long-time average of the DOS, $D_0^{\rm ave}$, in the AFM case.
As in the FM case in Fig.~\ref{fig_h_dep_dos}(a), $D_0^{\rm ave}$ increases while increasing $h$, but suddenly jumps to a larger value at $h_{c1}$.
We also find that $D_0^{\rm ave}$ shows a nonmonotonic change for $h_{c1}<h<h_{c2}$~\footnotemark[1].

In Fig.~\ref{fig_h_dep_dos}(b), we find that $D_0^{\rm ave}$ changes discontinuously at $h_{c1}$ despite the continuous transition in the equilibrium state [see Fig.~\ref{fig_hdep_full}(c)].
This suggests that the time evolution enhances the instability inherent to the equilibrium system and yields the large difference in the long-time behavior.
This is one of the significant features originating from the nonequilibrium dynamics.

\section{Discussion}\label{sec:discussion}

In this section, we discuss the results obtained in the previous section, with a focus on the possibilities of the experimental observation.
In the transient dynamics of the magnetization after the field quench from the Kitaev QSL, we find that the low-energy dynamics survives in the FM Kitaev model, while such behavior does not appear clearly for the quench from the low-field QSL in the AFM case after the high-energy dynamics dies out in the early stage.
For the quench from the intermediate QSL state in the AFM case, however, we obtain two kinds of dynamics with different lifetimes.
The typical time scale of these dynamics is $t/|J|^{-1}\sim10$--$100$.
It corresponds to 1--10~picosecond when we assume $|J|\sim 100$--$300$~K, which is expected in Kitaev candidate materials~\cite{PhysRevLett.110.097204,PhysRevB.88.035107,PhysRevLett.113.107201,1367-2630-16-1-013056,PhysRevB.91.241110,banerjee2016proximate}.
Although it might be difficult to control and measure the magnetization directly in experiments within this time scale, the optical techniques are likely applicable.
For instance, the optical control of the magnetization was achieved by using a laser pulse in the femtosecond order, via the inverse Faraday effect~\cite{kimel2005ultrafast,Stanciu2007}.
Moreover, the time evolution of the magnetization was observed by the time-resolved magneto-optical Faraday/Kerr effect with the picosecond order resolution~\cite{Hiebert1997,kimel2005ultrafast}.
Therefore, the present results may be observed by the optical measurements in the Kitaev candidate materials such as iridates and $\alpha$-RuCl$_3$, which enables us to examine the spin fractionalization in the different viewpoint from the equilibrium states and gives us the information on the lifetime of the quasiparticles.

In addition to the magnetization, the time evolution of the spin correlations could be measured by using the magnetic X-ray scattering.
In an equilibrium state, the spin correlations of each spin component were separately observed by the azimuthal angle dependence of the diffuse magnetic X-ray scattering, which evidenced the presence of bond-dependent interactions in the Kitaev candidate material Na$_2$IrO$_3$~\cite{chun2015direct}.
We expect that, if a similar experiment can be performed in the time domain, the time evolution of the spin correlations, especially $K$ which corresponds to the kinetic energy of the Majorana fermions $\{a,b\}$, may be measured in experiments.

Moreover, in the AFM case, we find the distinct transient dynamics between the low- and intermediate-field QSL phases as mentioned above.
By considering the fact that the topology of the Majorana fermion bands are different between these two phases~\cite{Nasu2018mag}, this suggests the possibility of differentiating the topological nature of the equilibrium states by the transient dynamics.
Unfortunately, it is difficult to test this interesting possibility for the available candidate materials, such as iridates and $\alpha$-RuCl$_3$, as the Kitaev interactions are predicted to be FM~\cite{PhysRevLett.113.107201,PhysRevB.93.174425,1367-2630-16-1-013056,PhysRevB.92.184411,Winter2016,Chaloupka2016,yadav2016kitaev}.
We note, however, that the AFM Kitaev interactions are theoretically proposed for $f$-electron based compounds~\cite{Jang2018pre}.

We also find the transient Fermi surfaces by the field quench.
In the equilibrium states at zero field, the Kitaev QSL on the honeycomb lattice does not have the Fermi surfaces~\footnote{The appearance of the Majorana Fermi surface was discussed in the presence of an effective staggered-type magnetic field~\cite{Takikawa2019pre}.}, while some extensions to three-dimensional lattices do~\cite{PhysRevB.89.235102}.
The appearance of the Fermi surfaces may cause the Peierls instability, as discussed for the equilibrium state in three dimensions~\cite{Hermanns_peierls2015}.
The Peierls instability appears stronger in lower dimensions.
Thus, our results suggest that the Peierls instability may appear in the transient dynamics, which never happens in equilibrium.
Although the energy dissipation is neglected in our calculations, if the dissipation time is longer than the typical time scale $t/|J|^{-1}\sim10$--$100$, such a Peierls instability associated with the transient Fermi surfaces is expected to occur.
It would lead to hidden phases, which cannot be reached in equilibrium, such as dimerized phases through the coupling to lattice deformations and symmetry-breaking phases by spontaneous Majorana ordering via quantum many-body effects.
In the present situation with the magnetic field along the $S^z$ direction, the transient Fermi surfaces are highly anisotropic and the resulting Peierls instability may occur along the $k_x$ direction, particularly for the quench from the intermediate state in the AFM case [see Fig.~\ref{fig_fermis}(c)].

\section{Concluding remarks}\label{sec:summary}

In summary, we investigated the nonequilibrium dynamics of the Kitaev model triggered by quenching the magnetic field.
Using the time-dependent Majorana MF theory, which takes into account the distinct energy scales of the fractional Majorana excitations, we examined the time evolutions of the magnetization and spin correlations.
We found that the spin fractionalization manifests itself as two kinds of transient dynamics in the case of the field quench from the Kitaev QSL state, which is distinct from the conventional spin precession in the case from the forced-FM one.
More peculiar two-stage dynamics is observed for the field quench from the intermediate QSL state in the AFM Kitaev model.
We discussed the origin of these peculiar dynamics from the energy transfer between the two types of Majorana fermions.
We also revealed the time evolution of the Majorana band structure and discussed the relation to the time evolutions of the magnetization and spin correlations.
In addition, we found the appearance of the transient ``Fermi surface'' and Majorana ``Lifshitz transition'', which cannot be achieved in static magnetic fields.
We proposed the possible observations of our results by the magneto-optical effects and the magnetic X-ray scattering.
We also discussed the possibility of emergent phases through the Peierls instability and quantum many-body effects between the Majorana fermions.

In the present study, we only addressed the magnetic field quench.
This is a first step toward a variety of intriguing real-time dynamics anticipated in QSLs with fractional excitations.
Our method can be straightforwardly extended to other time-dependent fields, such as pulse and ac magnetic fields.
Such extensions may reveal further peculiar transient dynamics related with the fractional excitations.

Although our approach was limited to the magnetic field along the $S^z$ direction because of the framework of the Majorana MF theory based on the Jordan-Wigner transformation, it could be extended to the time-dependent magnetic field along any direction by employing other MF approaches, such as the parton MF theory~\cite{Burnell2011,You2012,Schaffer2012,Knolle2018}.
Such an extension will pave the way for further theoretical and experimental studies on the transient dynamics of the Kitaev systems triggered by a greater variety of time-dependent fields, such as optical pumpings, spin pumping, and spin current injection.

\begin{acknowledgments}
The authors thank Y.~Kamiya, Y.~Kato, and J.~Ohara for helpful discussions.
This work is supported by Grant-in-Aid for Scientific Research under Grant No. JP16H02206, JP18H04223, and JP19K03742, and by JST CREST (JP-MJCR18T2).
Parts of the numerical calculations were performed in the supercomputing systems in ISSP, the University of Tokyo.
\end{acknowledgments}

\appendix

\section{Validity of time-dependent HF theory}\label{sec:valid-time-depend}

In this Appendix, we discuss the validity of the time-dependent Majorana MF theory introduced in Sec.~\ref{sec:time-dependent-hf} by calculating the dynamical spin structure factor in the limit of $h\to 0$ and comparing the results with the exact solutions.
Moreover, we also discuss the precision of the present method by examining the time evolution of the conserved quantities.

\subsection{Comparison of the dynamical spin structure factor at zero field with the exact solution}\label{sec:linear-response}

In the limit of $h\to 0$ in the field quench in Eq.~\eqref{eq:4}, $\Psi(t)=M^z(t)/h$ is nothing but the relaxation function.
Hence, by the linear-response theory, the dynamical magnetic susceptibility is calculated by
\begin{align}
 \chi(\omega)=\Psi(0^-)+i\omega \int_0^\infty e^{i(\omega+i\eta) t}\Psi(t) dt,\label{eq:5}
\end{align}
where $\eta$ is an infinitesimal real number.
As the dynamical spin structure factor at $\bm{q}=0$ is given by
\begin{align}
 {\cal S}(\bm{q}=0, \omega)=\frac{1}{N}\sum_{jj'}\int_{-\infty}^{\infty}\frac{dt}{2\pi}\means{S_j^z(t) S_{j'}^z}e^{i\omega t}=\frac{1}{\pi}{\rm Im} \chi,\label{eq:6}
\end{align}
we can obtain ${\cal S}(\bm{q}=0, \omega)$ at zero field by the time-dependent Majorana MF theory by taking $h\to 0$.

\begin{figure}[t]
 \begin{center}
  \includegraphics[width=\columnwidth,clip]{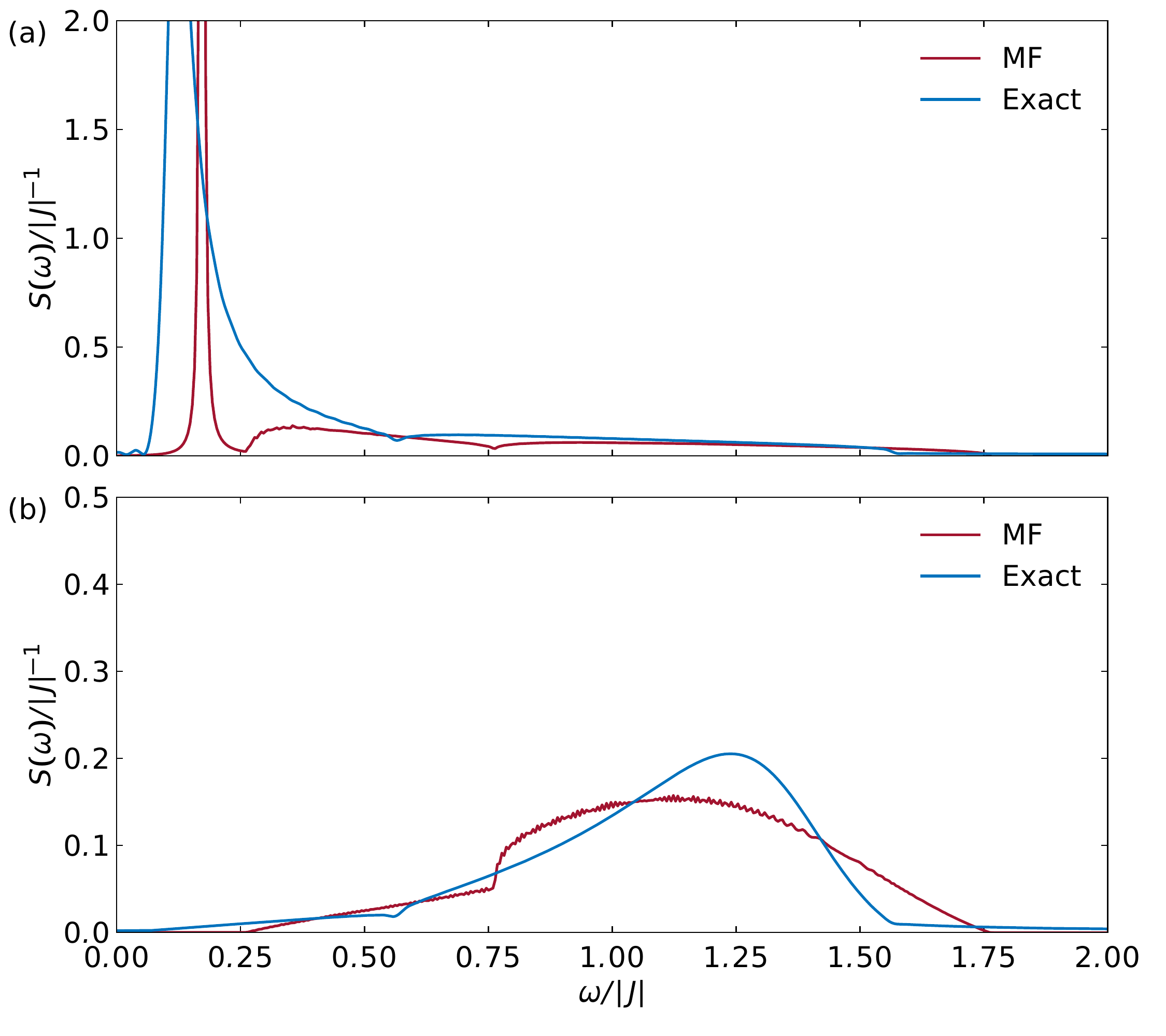}
  \caption{
Dynamical spin structure factor at $\bm{q}=0$, ${\cal S}(\bm{q}=0, \omega)$, for (a) the FM Kitaev model and (b) the AFM one.
The red lines indicate the results obtained by the Majorana MF theory and the blue lines are the exact solutions~\cite{PhysRevLett.112.207203,PhysRevB.92.115127}.
The exact results are calculated for $N=2\times 100^2$.
}
  \label{fig_sw}
 \end{center}
\end{figure}

Figure~\ref{fig_sw} displays the comparison between the time-dependent Majorana MF results and the exact solutions~\cite{PhysRevLett.112.207203,PhysRevB.92.115127}.
In the calculations, we take $h/|J|=0.0015$ and $\eta/|J|=0.00075$, and calculate the time evolution up to $t/|J|^{-1}=10^6$ with $\Delta t /|J|^{-1}=0.00667$.
In the FM case shown in Fig.~\ref{fig_sw}(a), the exact result exhibits the low-energy coherent peak and the high-energy broad structure, which predominantly originate from the localized and itinerant Majorana fermions, respectively.
The spectrum is zero in the low-energy part, reflecting the gap in the localized Majorana excitations.
In addition, in the broad structure, there is a small dip associated with the van-Hove singularity in the DOS of itinerant Majorana fermions.
All these characteristic features are qualitatively reproduced by the present time-dependent Majorana MF theory, although the low-energy coherent peak is considerably sharper than the exact one.
This is due to an overestimate of the lifetime of the localized Majorana fermions in the MF theory.
We note that our method also reproduces the low-energy spin gap behavior, which cannot be obtained by the classical-spin approach~\cite{Samarakoon2017}.
On the other hand, in the AFM case shown in Fig.~\ref{fig_sw}(b), the low-energy coherent peak is absent in the exact solution, leaving the broad spectrum in the wide energy range with a small dip due to the van-Hove singularity.
This is also reproduced by our Majorana MF theory.
These agreements suggest that the present calculations can capture the essential aspects of the time evolution of the fractional Majorana excitations in the Kitaev model in both FM and AFM cases.

We also examine the sum rule of the spectral weight:
\begin{align}
 \int_0^{\infty}{\cal S}(\omega)d\omega=\frac{1}{4}\pm \means{S_j^z S_{j'}^z}_z,
\end{align}
where $+$ ($-$) is for the FM (AFM) case.
The NN spin correlation on the right hand side is analytically evaluated~\cite{PhysRevLett.98.247201}.
We confirmed that our Majorana MF results satisfy the sum rule within good precision; the deviation is about $2\%$ ($0.5\%$) for the FM (AFM) Kitaev model.

\subsection{Conservation law}\label{sec:conservation-law}

\begin{figure}[b]
 \begin{center}
  \includegraphics[width=\columnwidth,clip]{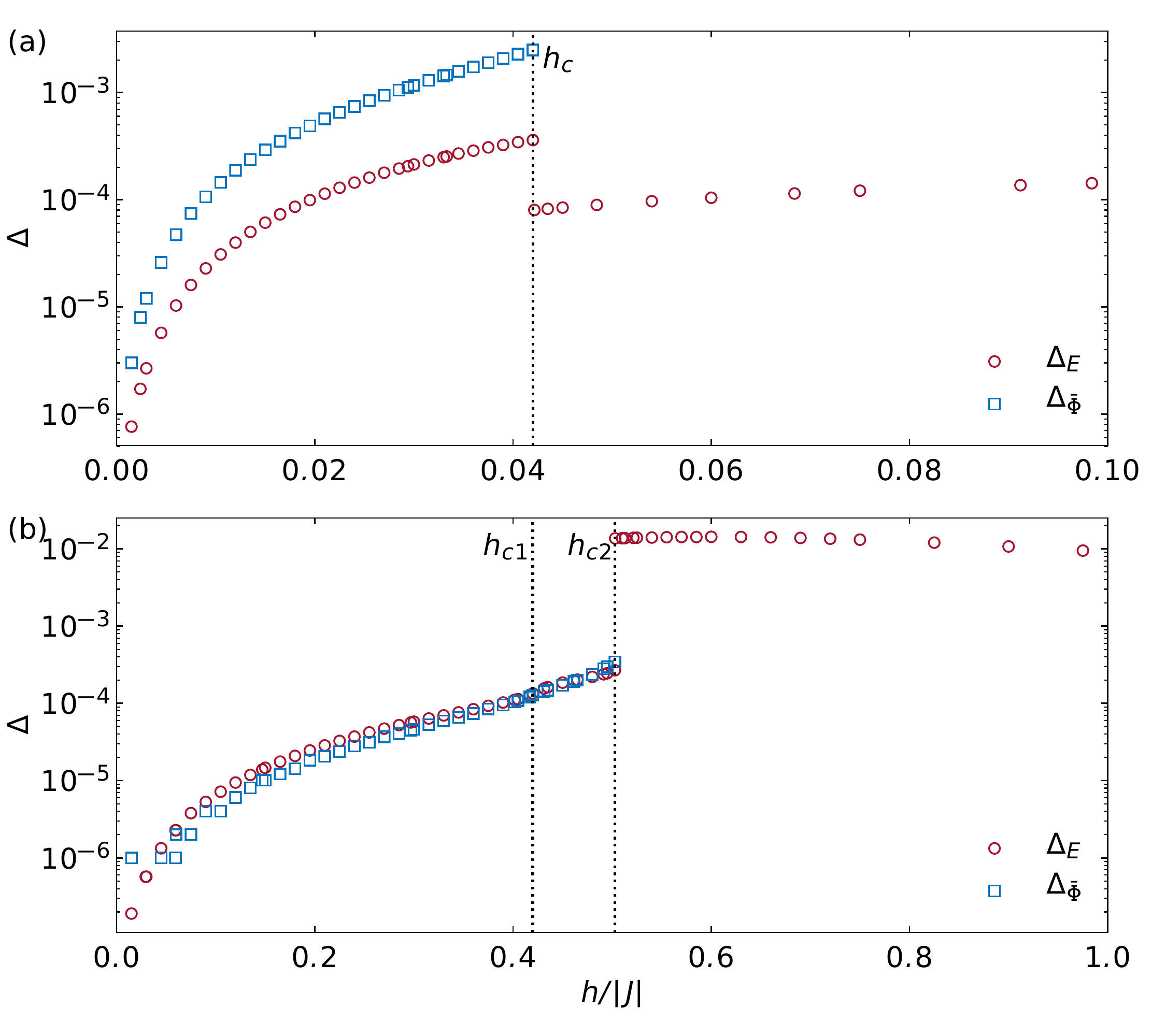}
  \caption{
Deviations from the conservation law for the total energy $E$ and the local conserved quantity $\bar{\Phi}(t)=i\means{\bar{a}_j\bar{b}_{j'}}$ in (a) the FM Kitaev model and (b) the AFM one.
See the text for the definitions.
}
  \label{fig_conservation}
 \end{center}
\end{figure}

To further test the validity of the time-dependent Majorana MF theory, we examine the time evolutions of the quantities that should be conserved.
In the present framework, in addition to the total energy $E$, the local quantity, $i\bar{a}_j\bar{b}_{j'}$ on the $z$ bond $\means{jj'}_z$, is also conserved.
We test the conservation of these quantities by calculating the deviation $\Delta_A=(A_{\rm max} -A_{\rm min})/A(0^-)$ for $A(t)=E(t)$ and $\bar{\Phi}(t)=i\means{\bar{a}_j\bar{b}_{j'}}$, where $A_{\rm max}$ ($A_{\rm min}$) is the maximum (minimum) value in the time range of $0<t/|J|<266.7$.
Figure~\ref{fig_conservation} shows the results.
Note that $\bar{\Phi}$ vanishes in the forced-FM phase, and hence, we do not show $\Delta_{\bar{\Phi}}$ above $h_c$ ($h_{c2}$) in the FM (AFM) case.
In the FM case, $\Delta_E$ and $\Delta_{\bar{\Phi}}$ increase while increasing $h$ in the range of $h<h_c$, but they remain smaller than $10^{-2}$, as presented in Fig.~\ref{fig_conservation}(a).
These quantities show similar $h$ dependence and even smaller in the AFM case below $h_{c2}$, as shown in Fig.~\ref{fig_conservation}(b).
However, above $h_{c2}$, $\Delta_E$ becomes larger than $10^{-2}$, which implies that the data in the forced-FM phase in the AFM case are not reliable in comparison with the other phases.

\bibliography{refs}

\end{document}